\documentstyle[12pt,equation,epsf]{article}
\def\epsffile#1{6.5truein}
\begin{document}
\newcommand{\cleqn}{\setcounter{equation}{0}}
\newcommand{\fac}{\frac{1}{8 \pi^2}}
\newcommand{\wino}{\mbox{$\widetilde{W}_1$}}
\newcommand{\zino}{\mbox{$\widetilde{Z}_1$}}
\newcommand{\mzi}{\mbox{$m_{\tilde{Z}_1}$}}
\newcommand{\mzit}{\mbox{$m_{\tilde{Z}_2}$}}
\newcommand{\mwi}{\mbox{$m_{\widetilde{W}_1}$}}
\newcommand{\mx}{\mbox{$M_X$}}
\newcommand{\muzero}{\mbox{$\mu_0^2$}}
\newcommand{\muone}{\mbox{$\mu_1^2$}}
\newcommand{\mutwo}{\mbox{$\mu_2^2$}}
\newcommand{\muthree}{\mbox{$\mu_3^2$}}
\newcommand{\mz}{\mbox{$m_Z$}}
\newcommand{\mt}{\mbox{$m_t$}}
\newcommand{\mb}{\mbox{$m_b$}}
\newcommand{\mg}{\mbox{$m_{\tilde g}$}}
\newcommand{\cg}{\mbox{$c_{\tilde g}$}}
\newcommand{\mpl}{\mbox{$M_P$}}
\newcommand{\sym}{\mbox{$SU(2)_L \times U(1)_Y$}}
\newcommand{\tanb}{\mbox{$\tan\!\beta$}}
\newcommand{\stw}{\mbox{$\sin^2\!\theta_W$}}
\newcommand{\sbo}{\mbox{$\tilde{b}_1$}}
\newcommand{\sto}{\mbox{$\tilde{t}_1$}}
\newcommand{\stot}{\mbox{$\tilde{t}_2$}}
\newcommand{\stau}{\mbox{$\widetilde{\tau}_1$}}
\newcommand{\msqav}{\mbox{$\langle m_{\tilde{q}} \rangle$}}
\newcommand{\kkbar}{\mbox{$K^0 - \overline{K^0}$}}
\newcommand{\gl}{\mbox{$\tilde{g}$}}
\newcommand{\mst}{\mbox{$m_{\tilde{t}_1}$}}
\newcommand{\mstt}{\mbox{$m_{\tilde{t}_2}$}}
\newcommand{\msul}{\mbox{$m_{\tilde{u}_L}$}}
\newcommand{\msb}{\mbox{$m_{\tilde{b}_1}$}}
\newcommand{\msq}{\mbox{$m_{\tilde{q}}$}}
\newcommand{\as}{\mbox{$\alpha_S$}}
\newcommand{\ttbar}{\mbox{$t \bar{t}$}}
\newcommand{\epem}{\mbox{$e^+ e^-$}}
\newcommand{\rs}{\mbox{$\sqrt{s}$}}
\newcommand{\mstau}{\mbox{$m_{\tilde{\tau}_1}$}}
\newcommand{\mhalf}{\mbox{$m_{1/2}$}}
\newcommand{\be}{\begin{equation}}
\newcommand{\ee}{\end{equation}}
\newcommand{\een}{\end{subequations}}
\newcommand{\ben}{\begin{subequations}}
\newcommand{\beq}{\begin{eqalignno}}
\newcommand{\eeq}{\end{eqalignno}}
\newcommand{\non}{\nonumber}
\newcommand{\wt}{\widetilde}
\newcommand{\ra}{\rightarrow}
\renewcommand{\thefootnote}{\fnsymbol{footnote} }
\pagestyle{empty}
\begin{flushright}
MAD--PH--879 \\
UM--TH--95--02\\
March 1995
\end{flushright}
\vspace*{1cm}
\begin{center}
{\Large \bf Implications of SUSY Model Building}\\
\vspace*{5mm}
 Manuel Drees\footnote{Heisenberg fellow} \\
{\it Dept. of Physics, Univ. of Wisconsin, Madison, WI 53706} \\
\vspace*{5mm}
and \\
\vspace*{5mm}
Stephen P. Martin \\
{\it Dept. of Physics, Univ. of Michigan, Ann Arbor, MI 48109--1120}
\end{center}
\begin{abstract}
\noindent
We discuss the motivations and implications of models of low-energy
supersymmetry. We present the case for the minimal supersymmetric
standard model, which we define to include the minimal particle content
and soft supersymmetry-breaking interactions which are universal at the
GUT or Planck scale. This model is in agreement with all present experimental
results, and yet depends on only a few unknown parameters and therefore
maintains considerable predictive power. From the theoretical side,
it arises naturally in the context of supergravity models. We discuss
radiative electroweak symmetry breaking and the superpartner spectrum in this
scenario, with some added emphasis on regions of parameter space leading to
unusual or interesting experimental signals at future colliders. We then
examine how these results may be affected by various modifications and
extensions of the minimal model, including GUT effects, extended gauge,
Higgs, and matter sectors, non-universal supersymmetry breaking,
non-conservation of R-parity, and dynamical supersymmetry breaking at
low energies.
\end{abstract}
\setcounter{page}{0}
\newpage
\pagestyle{plain}

\pagestyle{plain}
\setcounter{footnote}{0}
\section*{1) Introduction and Motivation}
``Supersymmetry is a symmetry that connects bosons and fermions." This
sentence, or something very much like it, used to be part of almost every talk
on supersymmetry until a few years ago. By now the concept of supersymmetry is
being taught in (graduate--level) classes, and speakers usually no longer
think it necessary to mention this elementary definition. However, for the
purpose of this article it might be helpful to keep in mind that ``testing
supersymmetry" means just that: To test whether there exists a (softly broken)
symmetry in nature that connects elementary bosons and fermions. It does {\em
not} mean testing a specific supersymmetric model, no matter how popular it
might be at a given time.

This is a noble goal. However, in practice it is impossible to discuss how the
existence of supersymmetry can be tested outside the context of some specific
realization of this symmetry. This realization might be described by a very
general ansatz with an abundance of free parameters, standing for ``a wide
class of models"; nevertheless it should be clear that we can never test {\em
all} softly broken SUSY models.\footnote{A falsifiable prediction of all
weakly coupled SUSY models is the existence of at least one light Higgs boson,
with mass less than 150 GeV \cite{101}. However, discovery of a light Higgs is
not quite an unambiguous signal for SUSY.} We therefore must focus on some
subset of models which we feel to be more reasonable, or in some sense more
``likely" to be a close approximation of the true theory. When making this
judgement it might be helpful to briefly recount the main motivations for the
introduction of SUSY.

Historically there have been two main arguments in favor of SUSY, one of which
might be called theoretical and the other very theoretical. The latter, purely
formal argument for space--time supersymmetry rests on the
Haag--Lopuszanski--Sohnius (HLS) theorem \cite{1}, which states that the
largest symmetry that a field theory can have without becoming trivial
(non--interacting) is the direct product of a (local) supersymmetry (the
algebra of which includes the generators of the Lorentz group) and an internal
gauge symmetry (with possibly very complicated group structure).

The description of particle physics in terms of relativistic field theories
has been very successful; by now it appears to be the only serious contender
for a realistic theory, at least for energies well below the Planck scale
$\mpl =2.4 \cdot 10^{18}$ GeV. Much of this success is due to the development
of gauge theories, which can describe all particle interactions we have seen
so far. Since we are doing a good job describing nature using two of the
three ingredients of the HLS theorem, isn't it likely that we'll do even
better once we incorporate the third ingredient -- supersymmetry -- as well?

This argument is strengthened further by the realization that, as remarked
above, the generators of the Lorentz algebra are a necessary ingredient of
the SUSY algebra if supersymmetry is to be realized locally. This allows one
to derive the theory of general relativity, our best classical theory of
gravity, from a deeper symmetry principle. A connection between SUSY and
gravity is also hinted at by the failure to develop a consistent quantum
theory of gravity that does {\em not} involve supersymmetry. The inconsistency
of all non--supersymmetric quantum gravity theories has not been proven
(other than ``by exhaustion"); nor does there exist a completely satisfying
supersymmetric quantum gravity theory. The fact remains, however, that all
recent attempts to quantize gravity perturbatively, including superstring
theories, involve supersymmetry \cite{2}.

This first argument says almost nothing about the scale where SUSY is broken.
Given that problems with quantum gravity seem to appear mostly at Planckian
energies, any SUSY breaking at or below \mpl\ would do. Of course,
superpartners (sparticles) with mass close to \mpl\ will not be produced in
laboratory experiments any time soon. Fortunately, the second main argument in
favor of supersymmetry points towards a much lower breaking scale.

This argument rests on the observation, most succinctly proven by Witten
\cite{3}, that the introduction of superpartners with mass not much above the
scale of electroweak symmetry breaking solves the technical part of the
``hierarchy problem" \cite{4}. The basic statement is that in supersymmetric
theories you are allowed to introduce widely separate energy scales, such as
the weak scale characterized by \mz\ and the scale \mx\ of grand unification
(GUT) or the Planck scale, without having to worry about quantum corrections
spoiling this hierarchy, as they tend to do in non--supersymmetric theories
with elementary scalars. In other words, in a nonsupersymmetric theory the
introduction of an elementary Higgs boson with mass of order \mz, rather than
${\cal O}(\mpl)$, is highly unnatural, while in a supersymmetric theory {\em
any} choice of the Higgs mass is technically natural, i.e. protected from
large quantum corrections, as long as the mass splitting between ``ordinary"
particles and their superpartners does not (greatly) exceed 1 TeV.

Of course, we do not know for sure that an elementary Higgs boson does indeed
exist; however, it remains the by far simplest, and theoretically most easily
treatable, realization of the idea of spontaneous breaking of gauge
symmetries. Indeed, while one can speak of supersymmetric theories, where
predictions for measurable quantities in terms of a few as--yet unknown
parameters can be derived reliably, the same cannot be said for models that
attempt to make do without elementary scalars. One might argue that this is
more a statement about our collective incompetence than about nature; however,
we feel that a theory that actually allows one to make predictions is
preferable to models where at best rough estimates are possible.

\makebox[15.5cm]{\epsfxsize=0.67\hsize\epsffile{sumo1.ps}}

\noindent {\bf Fig.~1:} {\small The number of papers on supersymmetry that
have been published in the last twenty years, according to the SLAC Spires
data base.} \\
\vspace*{4mm}

Witten's theorem \cite{3} only tells us that in a supersymmetric theory there
is nothing wrong with having the weak scale so much below the Planck scale; it
does not tell us how the weak scale got to be this small in the first place.
A glance at fig.~1 shows that SUSY only began to be widely taken seriously
after 1981/82, when a mechanism was discovered \cite{5} that might actually
explain why \mz\ is so small. Here one posits that at very high energies the
electroweak gauge symmetry is indeed unbroken; more formally, the scalar
(Higgs) potential renormalized at some scale close to the Planck or GUT
scale is assumed to have its minimum at the origin. Quantum corrections
drive the minimum away from the origin when the potential is probed at scales
of order \mz. These corrections are roughly $\propto \frac {\alpha} {4 \pi}
\log \frac {\mx}{\mz}$, where $\alpha$ is some generic coupling, and become
${\cal O}(1)$ only if $\mz \sim \mx \cdot e^{-4 \pi /\alpha}$ is exponentially
smaller than the more fundamental (input) scale \mx\ or \mpl. This for the
first time established a link between these widely separate scales. It has to
be admitted, however, that at present SUSY breaking masses very roughly of
order \mz\ still have to be introduced ``by hand"; the thorny issue of SUSY
breaking remains without definitive solution. While SUSY models do not
necessarily exhibit radiative gauge symmetry breaking, we feel that the
elegance of this mechanism justifies putting a strong emphasis in this article
on those models that do.

Fig.~1 shows a recent resurgence of SUSY's popularity, after the decline from
the peak years caused by the realization that the UA1 monojet events did not
signal the discovery of supersymmetry after all. This resurgence has been
triggered by refs.\cite{6}, which pointed out that LEP data clearly rule out
the unification of the $SU(3) \times SU(2) \times U(1)$ gauge group of the SM
into a GUT without the introduction of new particles and/or an intermediate
scale; moreover, the supersymmetric version of the SM {\em automatically}
contains just the right additional degrees of freedom to allow beautiful
unification of all gauge interactions, without the need for ad hoc
introduction of additional scales or exotic particles (other than those
dictated by supersymmetry, of course). The only assumption is that sparticle
masses do not greatly exceed 10 TeV. This is summarized in fig.~2, taken from
ref.\cite{7}. In our view this observation not only argues in favor of
weak--scale SUSY, but also argues {\em against} the existence of an
intermediate scale that significantly influences the running of the gauge
couplings, since such a scale would make the apparent unification shown in
fig.~2 a mere accident. The main emphasis of this article will therefore be on
models with a ``grand desert" between the weak and GUT scales.

SUSY models also tend to have other nice features. They allow \cite{8} for
unification of at least some Yukawa couplings within certain GUT models, which
might be a first step towards a complete understanding of the intricacies of
the quark and lepton spectrum. The introduction of superpartners also raises
the GUT scale sufficiently to make nucleon decay rates due to the exchange of
GUT gauge bosons fall safely below present experimental bounds. (SUSY GUTs
often have other, potentially more troublesome contributions to proton decay,
however \cite{9}.) Since sparticle mass terms in the Lagrangian respect all
gauge symmetries of the SM, sparticles decouple quickly from gauge
interactions once their mass exceeds the energy scale relevant for a given
experiment; in other words, quantum corrections due to sparticle loops tend to
be small. This is fortunate since electroweak precision experiments are in
good agreement with the SM. Finally, in many SUSY models the lightest
sparticle (LSP) is absolutely stable, and makes a good particle physics
candidate for the missing dark matter of the universe \cite{10}. This list of
at least qualitative successes becomes even more impressive when compared with
the track record of SUSY's main competitor, technicolor models \cite{11},
which have problems on many fronts, flavor changing neutral currents, the
electroweak $S$ parameter, and $Z \rightarrow b \bar{b}$ decays being the most
prominent ones.

\makebox[15.5cm]{\epsfxsize=0.53\hsize\epsffile{sumo2a.ps}}

\makebox[15.5cm]{\epsfxsize=0.53\hsize\epsffile{sumo2b.ps}}

\noindent {\bf Fig.~2:} {\small The running of the gauge couplings in the
Standard Model (top) and its supersymmetric extension (bottom). Both figures
assume $\alpha_S(\mz) = 0.120 \pm 0.01$. In the lower frame an effective SUSY
particle threshold at \mz\ has been assumed; adapted from ref.\cite{7}.} \\
\vspace{4mm}

The remainder of this article is organized as follows. Sec.~2 is devoted to
a rather thorough discussion of the minimal supersymmetric standard model
(MSSM), which we define to include radiative gauge symmetry breaking and a
simple sparticle spectrum at the GUT or Planck scale.\footnote{Our definition
differs from the one used in the chapter by Baer et al. \cite{53}.} This model
is fully realistic (in agreement with experiment), and yet has only a few free
parameters and hence considerable predictive power. After our definition of
``the MSSM" (2a) and a more detailed discussion of radiative symmetry breaking
(2b), we map out the parameter space in terms of physical masses in sec.~2c.
In sec.~2d we give handy analytical approximate solutions to the relevant
renormalization group equations, and sec.~2e discusses a few sample spectra in
more detail, with emphasis on those that give rise to somewhat unusual SUSY
signals at colliders. Finally, sec.~2f discusses the question whether useful,
strict upper bounds on sparticle masses can be given in the MSSM.

Sec.~3 treats minor modifications of the MSSM, where the ideas of a grand
desert and of radiative gauge symmetry breaking are kept but the details of
SUSY breaking are altered. In sec.~3a we discuss various sum rules that might
allow one to unravel such details, and in sec.~3b possible GUT effects on the
sparticle spectrum are summarized. Sec.~4 is devoted to more extensive
modifications of the MSSM. Sec.~4a deals with the question whether significant
statements about the sparticle spectrum can be made in grand desert models with
completely arbitrary SUSY breaking at the GUT scale, a scenario that has
recently been popularized in the string context. In secs.~4b, 4c and 4d models
with extended Higgs, gauge and matter sectors, respectively, are discussed.
Sec.~4e contains a brief summary of models with broken $R$ parity, and sec.~4f
is devoted to a discussion of recent attempts to break SUSY dynamically at
relatively low scales. Finally, in sec.~5 we make some concluding remarks.

\section*{2) The MSSM}
\subsection*{2a) Definition}
The minimal supersymmetric standard model (MSSM) is the by far most widely
studied potentially realistic SUSY model. It owes its popularity mostly to its
simplicity, being essentially a straightforward supersymmetrization of the SM.
However, different authors often make different assumptions about SUSY
breaking, which can lead to quite different phenomenological predictions. We
therefore start our discussion with a definition of the MSSM, to be used
throughout this article.

There is little disagreement about the particle content of the MSSM: It
contains three generations of (chiral) quark and lepton superfields, the
(vector) superfields necessary to gauge the $SU(3) \times \sym$ group of the
SM, and two (chiral) $SU(2)$ doublet Higgs superfields. The introduction of a
second Higgs doublet is necessary in order to cancel the anomalies produced by
the fermionic members of the first Higgs superfield, and also to give masses
to both charge $=+2/3$ and charge $=-1/3$ quarks. Notice that, while all matter
superfields contain elementary scalars, Higgs superfields still have to be
introduced separately; the boson--fermion symmetry therefore does not imply a
Higgs--matter symmetry.

The interactions between Higgs and matter superfields are described by the
superpotential
\be \label{e2.1}
W = h_U Q_L H_2 U_R + h_D Q_L H_1 D_R + h_E L_L H_1 E_R + \mu H_1 H_2.
\ee
Here $Q_L$ contains $SU(2)$ (s)quark doublets and $U_R$ and $D_R$ the
corresponding singlets; (s)lepton doublets and singlets reside in $L_L$ and
$E_R$, respectively, while $H_1$ and $H_2$ denote Higgs superfields with
hypercharge $Y = \mp 1/2$. In eq.(\ref{e2.1}) summation over gauge and
generational indices is implicit; in particular, $h_U, \ h_D$ and $h_E$ are
$3 \times 3$ matrices in generation space. Note that the only mass term in
(\ref{e2.1}) gives masses to Higgs bosons and fermions; the masses of all SM
fermions are due to the vacuum expectation values (vevs) $v_1 \equiv \langle
H_1^0 \rangle$ and $v_2 \equiv \langle H_2^0 \rangle$ that break \sym, in
close analogy to the SM. Since $\mu$ appears in the expressions for Higgs
masses even after SUSY breaking, it is clear that it must very roughly be of
order of the weak breaking scale; we saw in the Introduction that this must be
true for SUSY breaking masses as well. Yet within the MSSM $\mu$ is an
independent free parameter; it is not clear why it should be of the same order
of magnitude as sparticle masses. This is called the ``$\mu-$problem" of the
MSSM. One class of solutions \cite{71} makes use of non--renormalizable terms
either in the superpotential or in the K\"ahler potential appearing in the
supergravity Lagrangian; after SUSY breaking, $\mu$ is then automatically of
the order of the sparticle masses. Another possibility \cite{71a} is to
connect $\mu$ with the Peccei--Quinn solution \cite{71b} of the strong CP
problem, such that $\mu \sim M^2_{\rm PQ}/M_P$ with $M_{\rm PQ} \sim 10^{10}$
GeV, which again gives $\mu \sim 1$ TeV. Other solutions, which also connect
$\mu$ to the effective SUSY breaking scale, will be discussed in Secs.~3b1 and
4b.

If SUSY were exact, sfermions would have the same mass as the corresponding SM
fermions. This obviously not being the case in nature, we have to allow for
the presence of soft SUSY breaking terms when writing down the scalar
potential. Here ``soft breaking" means that the introduction of these terms
does not lead to the re--appearance of quadratically divergent quantum
corrections to scalar masses, the absence of which was one of the main
arguments in favor of SUSY. These terms have been classified in ref.\cite{12}
and include mass terms for sfermions and gauginos (but not for chiral
fermions) as well as trilinear and bilinear (but not quartic) scalar
interactions. The scalar potential can thus be written as:
\beq \label{e2.2}
V &= \sum_j \left| \frac {\partial W} {\partial \phi_j}\right|^2 + V_D \non \\
 &+ \sum_{i,j} m_{ij}^2 \phi_i^* \phi_j + A_U h_U \wt{Q}_L H_2 \wt{U}_R
+ A_D h_D \wt{Q}_L H_1 \wt{D}_R + A_E h_E \wt{L}_L H_1 \wt{E}_R + B \mu H_1
H_2,
\eeq
where $\phi_j$ stands for any scalar field, and the ``D-terms" $V_D$ can,
e.g., be found in ref.\cite{13}. The first line in (\ref{e2.2}) gives the
supersymmetric contributions to the potential, while the remaining terms break
SUSY.\footnote{In principle there can be additional soft breaking terms
\cite{hallran}, of the form $\phi_i \phi_j \phi^*_k$, e.g. $\wt{Q}_L H_1^*
\wt{U}_R, \ \wt{Q}_L H_2^* \wt{D}_R, \ \tilde{L}_L H_2^* \wt{E}_R$; however,
these terms do not arise in supergravity--type SUSY breaking scenarios, and
will lead to quadratic divergencies if there are chiral gauge singlet
superfields in the underlying theory.}

We have followed the usual convention in writing the trilinear soft breaking
terms as a product of a Yukawa coupling and an $A$ parameter. The products
$A_U h_U, \ A_D h_D$ and $A_E h_E$ are again $3 \times 3$ matrices; there
could also be flavor off--diagonal sfermion masses. In addition, we have to
introduce SUSY breaking masses $M_i, \ i=1,2,3$ for $U(1), \ SU(2)$ and
$SU(3)$ gauginos. Altogether one would then need more than 100 real parameters
to describe soft SUSY breaking in full generality. Clearly some simplifying
assumptions are necessary if we want to achieve something close to a complete
study, either theoretical or experimental, of parameter space.

In sec.~1 we presented arguments linking the electroweak scale with the Planck
or GUT scale across a grand desert. In addition, most recent attempts to
understand SUSY breaking connect it to Planck--scale physics in the form of
some supergravity or superstring theory \cite{2}. Finally, the ``unification
dogma" stipulates that physics ought to be in some sense simpler, i.e. more
symmetric, at very high energies. This leads us to the following ansatz for
the soft breaking parameters, which are viewed here as ``running" (scale
dependent) quantities:
\ben \label{e2.3} \beq
M_i(\mx) &= \mhalf, \ i=1,2,3 ; \label{e2.3a} \\
m^2_j(\mx) &= m_0^2, \ j=\tilde{u}, \tilde{d}, \dots; \label{e2.3b} \\
A_k(\mx) &= A, \ k=U,D,E. \label{e2.3c}
\eeq \een
This reduces the number of free parameters describing SUSY breaking to just
four: The gaugino mass \mhalf, the scalar mass $m_0$, and the trilinear and
bilinear soft breaking parameters $A$ and $B$. We do not assume any relation
between $A$ and $B$, e.g. $A = B + m_0$, since such relations tend to be
very model--dependent \cite{14}. We do assume unification of the gauge
couplings at scale $\mx \simeq 2 \cdot 10^{16}$ GeV, but fig.~2 shows that
this is more or less automatic anyway as long as $m_0$ and \mhalf\ do not
exceed 10 TeV or so. At this point we do not assume any relations between
Yukawa couplings, since they are again very dependent on the specific GUT
model.\footnote{From a theoretical perspective it would be more natural to
impose boundary conditions like eqs.(\ref{e2.3}) at the Planck or perhaps
string compactification scale $M_C \simeq 5 \cdot 10^{17}$ GeV, rather than
the GUT scale. In string theory unification of the gauge couplings can be
achieved even in the absence of a GUT; in this case it makes little difference
whether we impose eqs.(\ref{e2.3}) at scale \mx\ or $M_C$. On the other hand,
if a GUT does exist, and some of its superheavy fields have large Yukawa
couplings, weak--scale predictions can be affected substantially. We will come
back to this point in sec.~3b.}

The values of the soft breaking parameters at experimentally accessible
energies can be computed from a set of coupled renormalization group equations
(RGE), which were first derived by Inoue et al.\cite{15}.\footnote{Recently
these equations have been extended to two--loop order \cite{15a}. No numerical
study of the sparticle spectrum using these refined equations has yet been
published (see, however, ref.\cite{15b}). Generally one expects two--loop
effects to be comparable in size to one--loop threshold corrections, which can
be computed only once the GUT (or string) model has been specified.} This
leads us to radiative gauge symmetry breaking, which is the topic of the next
subsection.

\setcounter{footnote}{0}
\subsection*{2b) Radiative Symmetry Breaking}
As remarked in sec.~1, the possibility to break the electroweak \sym\
symmetry of the SM radiatively is widely regarded as one of the main arguments
in favor of SUSY, since it offers a dynamical explanation for the mysterious
negative squared mass of the Higgs boson, which has to be introduced ``by
hand" in the SM. In fact, in the MSSM as defined in the previous subsection,
radiative symmetry breaking is a {\em necessity}. Understanding radiative
symmetry breaking is therefore essential for a full appreciation of the
MSSM.

Focusing on the neutral components of the Higgs doublets, the scalar
potential (\ref{e2.2}) reads \cite{16}:
\be \label{e2.4}
V_0^{\rm Higgs} = m_1^2 |H_1^0|^2 + m_2^2 |H_2^0|^2 + B \mu \left( H_1^0
H_2^0 + h.c. \right) + \frac{g_1^2+g_2^2}{8} \left( |H_1^0|^2 - |H_2^0|^2
\right)^2,
\ee
where $g_1$ and $g_2$ are the $U(1)$ and $SU(2)$ gauge couplings and the
subscript 0 in $V_0^{\rm Higgs}$ indicates that for the moment we are only
interested in the tree--level (classical) potential, although the parameters
in eq.(\ref{e2.4}) are scale--dependent (running). Notice that the quartic
term vanishes in the direction $|H_1^0| = |H_2^0|$; the potential will
therefore only be well--behaved (bounded from below) if
\be \label{e2.5}
m_1^2 + m_2^2 > 2 |B \mu|.
\ee
Further, when trying to minimize $V_0^{\rm Higgs}$ one finds a minimum with
nonvanishing vevs only if
\be \label{e2.6}
m_1^2 m_2^2 < |B \mu|^2.
\ee
On the other hand, the boundary conditions (\ref{e2.3}) imply that
\be \label{e2.7}
m_1^2(\mx) = m_2^2(\mx) = m_0^2 + \mu^2(\mx).
\ee
It is quite easy to convince oneself that conditions (\ref{e2.5}) and
(\ref{e2.6}) {\em cannot} be fulfilled simultaneously if $m_1^2=m_2^2$,
thereby seemingly excluding the possibility of consistent \sym\ breaking.

The solution of this dilemma rests on the observation \cite{5,15} that
quantum corrections lift the degeneracy of the Higgs masses. Eq.(\ref{e2.7})
only holds at the GUT scale; the values of $m_1^2$ and $m_2^2$ at lower
scales can be computed from the RGE given in ref.\cite{15}. It is easy to
understand that these two Higgs masses will run differently. A glance at the
superpotential (\ref{e2.1}) shows that $H_2$ couples to top (s)quarks, while
$H_1$ only couples to $b$ (s)quarks and $\tau$ (s)leptons. These couplings
are expected to be of very different strengths, since the top quark is much
heavier than all other SM fermions. The crucial observation is that Yukawa
couplings enter the RGE for squared scalar masses with positive sign; since we
are running the RGE ``backward", from \mx\ down to the weak scale, the Yukawa
terms tend to {\em reduce} the squared Higgs masses, until eventually
condition (\ref{e2.6}) is satisfied and the gauge symmetry is broken. Since
for scales $Q < \mx$ one has $m_1^2 > m_2^2$ it is generally not difficult to
choose parameters such that inequality (\ref{e2.5}) remains true at all
scales. Finally, while the top Yukawa coupling also influences the running of
the squared stop masses, due to Casimir factors appearing in the RGE \cite{15}
$m_2^2$ is reduced more quickly than $m^2_{\tilde{t}_{L,R}}$; again, it is not
difficult to make sure that squared stop masses remain positive even when
condition (\ref{e2.6}) is satisfied.

This is demonstrated in fig.~3, where we plot $m^2_{H_{1,2}}$ (defined as
$m^2_{1,2} = m^2_{H_{1,2}} + \mu^2$ at all scales) and $m^2_{\tilde{t}_R}$ as
a function of the scale $Q$, for two different sets of parameters leading to a
moderate and a large value for the ratio of vevs $\tanb = v_2/v_1$. In both
cases $m^2_{H_2}$ is decreased most quickly, and $m^2_{\tilde{t}_R}$ actually
increases with decreasing $Q$ due to gluino loop contributions \cite{15}.
($m^2_{\tilde{t}_L}$ grows even faster.) For small or moderate \tanb,
$m^2_{H_1}$ also increases with decreasing $Q$ or remains more or less
constant, whereas for large \tanb\ it decreases, albeit more slowly than
$m^2_{H_2}$ does. This dependence on \tanb\ follows from the relation between
Yukawa couplings and quark masses:
\ben \label{e2.8} \beq
h_b = \frac{m_b}{v_1} &= \frac {g_2 m_b}{\sqrt{2} \cos \! \beta m_W}
\rightarrow \frac {g_2 m_b}{\sqrt{2} m_W} \tanb; \label{e2.8a} \\
h_t = \frac{m_t}{v_2} &= \frac {g_2 m_t}{\sqrt{2} \sin \! \beta m_W}
\rightarrow \frac {g_2 m_t}{\sqrt{2} m_W}, \label{e2.8b}
\eeq \een
where $\rightarrow$ refers to the limit $\tan^2 \beta \gg 1$. Large \tanb\
therefore implies a large bottom Yukawa coupling, which leads to a sizable
reduction of $m^2_{H_1}$ when going from \mx\ down to the weak
scale.\footnote{Note that for the case $\tanb=5$ shown in fig.~3, $m_2^2 =
m^2_{H_2} + \mu^2$ remains positive at all scales, but condition (\ref{e2.6})
is nevertheless fulfilled. Similarly, even though $m^2_{H_1}$ eventually
turns negative for $\tanb=35$, $m_1^2$ remains well above zero and condition
(\ref{e2.5}) remains satisfied.}

\makebox[15.5cm]{\epsfxsize=0.67\hsize\epsffile{sumo3.ps}}

\noindent {\bf Fig.~3:} {\small The running of the SUSY breaking squared Higgs
and $\tilde{t}_R$ masses {}from $\mx = 2 \cdot 10^{16}$ GeV to the weak
scale, for a running top mass $\mt(\mt)=165$ GeV and two different values of
\tanb.}\\
\vspace*{4mm}

Of course, we do not merely want \sym\ to be broken; we also want the $Z$
boson to have the experimentally measured mass. Furthermore, in view of the
strong dependence of some weak--scale quantities on \tanb\ shown in fig.~3,
it is often more convenient to treat it as an independent input parameter.
Since $m_Z^2 = \frac {g_1^2 + g_2^2}{2} \left( v_1^2 + v_2^2 \right)$, fixing
\mz\ and \tanb\ determines both vevs $v_1$ and $v_2$. On the other hand, these
vevs can also be computed by minimizing the potential (\ref{e2.4}). Requiring
$\frac {\partial V_0^{\rm Higgs} } {\partial v_1} = \frac {\partial
V_0^{\rm Higgs} } {\partial v_2} = 0$ then gives:
\ben \label{e2.9} \beq
\mu^2 &= \frac {m^2_{H_2} \sin^2 \beta - m^2_{H_1} \cos^2 \beta} {\cos \! 2
\beta} - \frac{1}{2} m_Z^2; \label{e2.9a} \\
2 B \mu &= \tan \! 2 \beta \left( m^2_{H_1} - m^2_{H_2} \right)
+ m_Z^2 \sin \! 2 \beta. \label{e2.9b}
\eeq \een
Recall that $h_t > h_b$ implies $m^2_2 < m^2_1$ and thus $v_2 > v_1$, which
means that $\cos \! 2 \beta < 0$; it is then easy to see that $\mu^2 > 0$ in
the cases depicted in fig.~3. On the other hand, eq.(\ref{e2.9b}) only allows
to fix the sign of $B \mu$. Since the $B$ parameter appears nowhere else in
the theory (unlike $\mu$, which appears in chargino, neutralino and sfermion
mass matrices), we can satisfy eq.(\ref{e2.9b}) for either sign of $\mu$ by
choosing the sign of $B$ appropriately; in general we therefore have to
consider both signs of $\mu$. It is important to notice, however, that in this
scheme the value of $|\mu|$ is {\em fixed} by our choice of $m_0, \ \mhalf, \
A, \ \tanb$ and \mt, all of which affect the values of $m^2_{H_{1,2}}$ at the
weak scale.

The Yukawa couplings are the driving force of electroweak symmetry breaking
in the MSSM. Fortunately we now have good evidence \cite{17} that the mass of
the top quark lies in the vicinity of 175 GeV. For most values of \tanb\ and
the soft breaking parameters, varying \mt\ by about $2\sigma$ to either side
of this central value does not alter the predictions for the spectrum
dramatically, although the value of $|\mu|$ computed from eq.(\ref{e2.9a}),
and of the masses of the heavier physical Higgs bosons of the model \cite{16},
still do vary significantly. This is illustrated in fig.~4a, where we show
predictions for the masses of the lighter stop and stau eigenstates, for the
pseudoscalar Higgs mass $m_P$, and for $\mu$ at the weak scale, as a function
of the running top mass $m_t(m_t)$, which is the mass appearing in
eq.(\ref{e2.8b}); it is related to the physical (on--shell) top mass by
\cite{18}
\be \label{e2.10}
m_t^{\rm ON} = m_t (m_t) \left[ 1 + \frac{4}{3} \frac{\alpha_s(m_t)}{\pi}
 + 11.0 \left( \frac {\alpha_s(m_t)} {\pi} \right)^2 \right],
\ee
where we have ignored smaller electroweak corrections. The CDF central value
\cite{17} $m_t^{\rm ON}=174$ GeV corresponds to $m_t(m_t) \simeq 165$ GeV,
which we will adopt as out standard choice from now on.

\makebox[15.5cm]{\epsfxsize=0.52\hsize\epsffile{sumo4a.ps}}

\makebox[15.5cm]{\epsfxsize=0.52\hsize\epsffile{sumo4b.ps}}

\noindent {\bf Fig.~4:} {\small The dependence of the physical \sto, $P$ and
\stau\ masses as well as the value of $|\mu|$ at the weak scale as a function
of the running top mass (a) and \tanb\ (b). The value of $|\mu|$ has been
fixed by the requirement of correct electroweak symmetry breaking,
eq.(\ref{e2.9a}). Notice that in (b), $m^2_P$ turns negative at $\tanb \simeq
51$, which determines the upper bound on this quantity for the given choice of
parameters.}\\

The qualitative trends in fig.~4a are quite easy to understand. For given SUSY
breaking parameters, increasing \mt\ will decrease $m^2_{H_2}$ leaving
$m^2_{H_1}$ more or less the same, thereby increasing $|\mu|$ as dictated by
eq.(\ref{e2.9a}). At tree level, the mass of the pseudoscalar Higgs boson is
simply given by
\be \label{e2.11}
m^2_P = m_1^2 + m_2^2 = m^2_{H_1} +  m^2_{H_2} + 2 \mu^2;
\ee
an increase of $|\mu|$ therefore leads to an increase of $m_P$ as well.

The masses of the lighter stop and stau eigenstates can be computed from the
following mass matrices \cite{19} (we follow the sign conventions of
ref.\cite{20}):
\ben \label{e2.12} \beq
{\cal M}^2_{\tilde{t}} &= \mbox{$
\left( \begin{array}{cc} m^2_{\tilde{t}_L} + m^2_t
+ 0.35 D & -\mt ( A_t + \mu \cot \! \beta ) \\
-\mt ( A_t + \mu \cot \! \beta) & m^2_{\tilde{t}_R} + m^2_t + 0.16 D
\end{array} \right)$}; \label{e2.12a} \\
{\cal M}^2_{\tilde{\tau}} &= \mbox{$
\left( \begin{array}{cc} m^2_{\tilde{\tau}_L} +
m^2_{\tau} - 0.27 D & -m_{\tau} ( A_{\tau} + \mu \, \tanb ) \\
-m_{\tau} ( A_{\tau} + \mu \, \tanb) & m^2_{\tilde{\tau}_R} + m^2_{\tau} -0.23D
\end{array} \right)$}, \label{e2.12b}
\eeq \een
where $D = m_Z^2 \cos \! 2 \beta < 0$. \mst\ decreases with increasing \mt\
since the top Yukawa coupling decreases $m^2_{\tilde{t}_{L,R}}$ compared to
other squark masses. \mstau\ here depends very little on \mt\ since for the
given choice of \tanb, $\wt{\tau}_L - \wt{\tau}_R$ mixing is still negligible.

Fig.~4b shows that the dependence of the same masses on \tanb\ tends to be
be stronger than on \mt, at least when \tanb\ is varied over its entire
allowed range. The lower end of this range is determined by the requirement
that the top Yukawa coupling $h_t$ remains finite all the way up to the GUT
scale. This implies \cite{8}
\be \label{e2.13}
\sin \! \beta \geq \frac {m_t(m_t)} {192 \ {\rm GeV}},
\ee
or $\tanb \geq 1.65$ for $m_t(m_t) = 165$ GeV.

As \tanb\ is increased from this minimal value, the top Yukawa coupling
initially decreases quite rapidly, see eq.(\ref{e2.8b}); moreover, $|\cos \! 2
\beta|$ increases. Both effects reduce $|\mu|$ as computed from
eq.(\ref{e2.9a}). The decrease of $h_t$ and $|\mu|$ leads to the observed
increase of \mst. However, for $\tanb \geq 5$, $\mu$ becomes more or less
independent of \tanb\ since now $\sin \! \beta \simeq 1$ and $\cos \! 2 \beta
\simeq -1$ anyway. At the same time the bottom Yukawa coupling keeps
increasing, and hence $m^2_{H_1}$ at the weak scale keeps decreasing.
Eventually this leads to $m_P^2 <0$, i.e. an unstable Higgs potential
(condition (\ref{e2.5}) is violated). In the absence of hypercharge
interactions and for vanishing $h_{\tau}$, this happens for $\tanb =
m_t(m_t)/m_b(m_t)$ where $h_t=h_b$ so that the equality $m_1^2 = m_2^2$ again
holds at all scales. The small effects from hypercharge interactions and the
$\tau$ Yukawa coupling on this upper bound tend to cancel, so that one has to
good approximation
\be \label{e2.14}
\tanb \leq \frac {m_t(m_t)} {m_b(m_t)}.
\ee
Notice that the running bottom mass at scale \mt\ is relevant here, which
amounts to about 3 GeV if $m_b(m_b) \simeq 4.5$ GeV. If \tanb\ is close to
this upper bound, \mstau\ is also reduced significantly, due to the decrease
of the diagonal masses $m^2_{\tilde{\tau}_{L,R}}$ and the increase of the
off--diagonal entries in the mass matrix (\ref{e2.12b}). The mass of the
light sbottom eigenstate behaves very similarly \cite{20}; however, while
\stau\ is dominantly an $SU(2)$ singlet, \sbo\ is mostly doublet since
$m_{\tilde{b}_L}$ is affected by the dominant top Yukawa coupling while
$m_{\tilde{b}_R}$ only feels the influence of $h_b$.

So far we have limited our discussion to the RGE--improved tree--level
potential. As pointed out in ref.\cite{22}, in general one can greatly improve
the reliability of the calculation by including at least the leading terms
of the 1--loop corrections to the potential, given by \cite{23}
\be \label{e2.15}
V_1^{\rm Higgs} = \frac {1} {64 \pi^2} Str {\cal M}^4(H_1,H_2) \left[
\log \frac { {\cal M}^2(H_1,H_2) } {Q_0^2} - \frac {3}{2} \right],
\ee
where bosons (fermions) contribute with positive (negative) sign in the
supertrace, and the mass matrix ${\cal M}$ has to be considered to be a
function of the Higgs fields $H_1$ and $H_2$. The scale $Q_0$ in
eq.(\ref{e2.15}) is to be identified with the scale where the RG running of
the parameters in $V_0^{\rm Higgs}$ is terminated. The sum $ V_0^{\rm Higgs}
+ V_1^{\rm Higgs}$ then depends only weakly on $Q_0$, in sharp contrast with
$V_0^{\rm Higgs}$ by itself, the remaining scale dependence being due to
wave--function renormalizations and higher loops, both of which are small
effects. If one is satisfied with 5\% accuracy it is generally sufficient to
only include (s)top and perhaps (s)bottom contributions to the supertrace.

Notice that the 1--loop correction (\ref{e2.15}) is linear in $\log Q_0^2$; it
is therefore always possible to find a value of $Q_0$ such that this
correction is quite small in the minimum of the potential. In many cases it is
sufficient to treat the minimization of the potential by simply using the
tree--level relations (\ref{e2.9}) with $Q_0^2 \simeq m_0^2 + 3 m^2_{1/2}$,
which is something like an average squared stop mass; the same choice also
ensures that corrections to the masses of the pseudoscalar Higgs boson
(\ref{e2.11}), as well as the charged Higgs, are quite small. However, at
least the 1--loop corrections from the (s)top sector should always be included
when calculating the masses of the physical scalar Higgs bosons, in particular
the lighter one \cite{26,24,25}; in the MSSM they increase the upper bound on
its mass from \mz\ to something like 135 GeV.\footnote{The scalar Higgs masses
can be computed from the second derivatives of the Higgs potential at its
minimum. One finds that the entire $\log Q_0^2$ dependence of these
derivatives can be absorbed in the mass of the pseudoscalar Higgs boson,
leaving a large, $Q_0$ independent 1--loop correction to the scalar Higgs
masses behind \cite{24}.}

This completes our discussion of the mechanism of radiative symmetry breaking.
We are now ready to explore the parameter space of the MSSM in a more
systematic way.

\setcounter{footnote}{0}
\subsection*{2c) The Particle Spectrum of the MSSM}
In this subsection we explore the parameter space of the MSSM, with emphasis
on the resulting particle spectrum. The masses of squarks and sleptons of the
first two generations, whose Yukawa couplings are too small to affect the
running of their masses significantly, can be computed from the input
parameters $m_0, \ \mhalf$ and \tanb\ using the following well--known
expressions \cite{13}:
\ben \label{e2.16} \beq
m^2_{\tilde{e}_R} &= m_0^2 + 0.15 m^2_{1/2} - \stw D; \label{e2.16a} \\
m^2_{\tilde{e}_L} &= m_0^2 + 0.52 m^2_{1/2} - \left(\frac{1}{2} -
\stw \right) D; \label{e2.16b} \\
m^2_{\tilde{\nu}} &= m_0^2 + 0.52 m^2_{1/2} + \frac{1}{2}D; \label{e2.16c} \\
m^2_{\tilde{u}_R} &= m_0^2 + (0.07 + \cg) m^2_{1/2} + \frac{2}{3} \stw D;
 \label{e2.16d} \\
m^2_{\tilde{d}_R} &= m_0^2 + (0.02 + \cg) m^2_{1/2} - \frac{1}{3} \stw D;
 \label{e2.16e} \\
m^2_{\tilde{u}_L} &= m_0^2 + (0.47 + \cg) m^2_{1/2} + \left(\frac{1}{2}-
\frac{2}{3} \stw \right) D;  \label{e2.16f} \\
m^2_{\tilde{d}_L} &= m_0^2 + (0.47 + \cg) m^2_{1/2} - \left(\frac{1}{2} -
\frac{1}{3} \stw \right) D, \label{e2.16g}
\eeq \een
where $D = m_Z^2 \cos \! 2 \beta$ as before. The coefficient \cg\ describes
the contribution from gluino--quark loops to the running of squark masses and
is given by
\be \label{e2.17}
\cg = \frac{8}{9} \left[ \left( \frac {\as(\msq)} {\as(\mx)} \right)^2 - 1
\right].
\ee
Notice that \cg\ depends logarithmically on the squark mass in question.
However, it is usually sufficient to take $m^2_{\tilde q} = m_0^2 + 6
m^2_{1/2}$ on the r.h.s. of eq.(\ref{e2.17}). Numerically \cg\ varies between
approximately 4.5 and 6 for squark masses between 0.1 and 1
TeV.\footnote{Eqs.(\ref{e2.16d})--(\ref{e2.16g}) compute the running squark
masses. Eq.(\ref{e2.17}) makes sure that they are computed at scale
$Q=\msq$; the difference between running and on--shell mass is then quite
small.}

The values of the gaugino masses at the weak scale are also easily computed
\cite{13}:
\be \label{e2.18}
M_i(Q) = \mhalf \frac {\alpha_i(Q)}{\alpha_i(\mx)}.
\ee
Numerically, $M_1 \simeq 0.41 \mhalf$ and $M_2 \simeq 0.84 \mhalf$ for $Q
\simeq m_Z$; the $Q-$dependence is quite mild here. This is not true for the
running gluino mass $M_3$, however. In addition, the difference between the
physical (on--shell) and running mass can be quite significant in this
case \cite{27}:
\be \label{e2.19}
\mg = M_3(Q) \left\{ 1 + \frac {\as(Q)}{4 \pi} \left[ 15 - 18 \log \frac
{M_3(Q)} {Q} + \sum_{\tilde q} B_1(M_3^2,\msq,m_q) \right] \right\},
\ee
where $B_1$ is the (finite part of) a standard 2--point function:
\beq \label{e2.20}
B_1(s,m_1,m_2) &= \int_0^1 dx x \log \frac {x m_1^2 + (1-x) m_2^2 - x (1-x) s}
{Q^2} \non \\
&\simeq \log \frac{m_1}{Q} \ \ \ (m_1^2 \gg m_2^2,s).
\eeq
The difference between \mg\ and $M_3(M_3)$ can amount to 30\% if $M_3 \simeq
0.1$ TeV and $\msq \simeq 1$ TeV. Notice that the physical gluino mass \mg\
in (\ref{e2.19}) is (to 1--loop order) independent of the scale $Q$.

The masses of third generation sfermions and Higgs bosons, as well as the
$\mu$ parameter, are slightly more difficult to compute. If only the top
Yukawa coupling is significant, expressions can be given that only depend on
a single integral that has to be computed numerically \cite{28}. If $h_b$
and/or $h_{\tau}$ are non--negligible, i.e. for large \tanb, no such
semi--analytical solutions can be found. In this subsection we will use fully
numerical methods to integrate the RGE, implement radiative symmetry breaking,
and compute the (s)particle spectrum at the weak scale. By now many such
programs have been written \cite{5,28,20,29,44}.

Since even for fixed \mt\ the MSSM contains four free parameters ($m_0, \
\mhalf, \ A$ and \tanb), the mass of any one particle can usually vary from
its experimentally defined lower limit to (effectively) infinite; the only
(very important) exception is the mass of light Higgs boson, which cannot
exceed $\sim 135$ GeV in the MSSM, as already noted. In contrast, there are
often quite striking correlations between {\em pairs} of masses; in other
words, it is often not possible to choose two masses independently from each
other. A well--known example \cite{30} is given by the gluino mass and the
first or second generation squark masses, collectively denoted by \msq:
\be \label{e2.21}
\msq \geq 0.85 \mg,
\ee
which follows from eqs.(\ref{e2.16})--(\ref{e2.19}) and the requirement
$m_0^2 \geq 0$. This inequality means that one of the two often studied
extreme cases of squark and gluino searches at hadron colliders \cite{31},
viz. $\msq \ll \mg$, cannot be realized in the MSSM.\footnote{The CDF study
tries to incorporate part of the MSSM by linking the LSP and gluino masses
via eq.(\ref{e2.18}). This leads to the somewhat bizarre statement that
scenarios with $\mg \gg \msq$ cannot be ruled out experimentally since the
LSP becomes so heavy that there is little missing $E_T$ in the event.
Including eq.(\ref{e2.18}) but ignoring the constraint (\ref{e2.21}) offers
a prime example how model predictions should {\em not} be used in experimental
analyses.}

Other correlations are slightly less straightforward to derive since they
involve radiative symmetry breaking. Examples are shown in figs.~5a--j, where
we show the allowed regions of the planes spanned by the parameters $M_2$ and
$\mu$ appearing in chargino and neutralino mass matrices (fig.~5a), as well as
by pairs of the physical masses $m_P, \ \mst, \ \mstau, \ \msul, \ \mzi$ and
\mwi; here \zino\ and \wino\ denote the lightest neutralino and chargino,
respectively, and $\tilde{u}_L$ has been chosen as a ``typical" first or
second generation squark.

We have fixed $\mt(\mt)=165$ GeV and the GUT gauge coupling $g_X = 0.71$ at
$M_X = 1.9 \cdot 10^{16}$ GeV\footnote{For consistency we use 1--loop RGE
everywhere}; we have considered three different values of \tanb. $\tanb=1.65$
(long dashed) corresponds to the case of maximal top quark Yukawa coupling
(so--called fixed point scenario \cite{21}, since this maximal coupling at the
weak scale is approached from a wide range of choices for $h_t(\mx)$; in
practice we have chosen $h_t(\mx)=2$.) This choice is of interest not only
because maximizing $h_t$ obviously maximizes its contributions to the RGE, but
also because it allows \cite{8} the $b$ and $\tau$ Yukawa couplings to be
unified, $h_b(\mx)=h_{\tau}(\mx)$. Moreover, large Yukawa couplings are
preferred dynamically \cite{45a} in certain superstring models where these
couplings are determined by MSSM quantum effects. The case $\tanb=8$ (solid)
stands for the ``typical" situation where $h_t$ is already almost independent
of \tanb\ (with $h_t(\mx) \simeq 0.53$), but $h_b$ and $h_{\tau}$ are still
essentially negligible in the RGE. This is no longer true for our third
choice, $\tanb=35$ (short dashed), although this is still well below the
maximal value of \tanb, see fig.~4b.

The allowed region of parameter space is defined by the following constraints:
\begin{itemize}
\item For the given (large) value of \mt, radiative symmetry breaking always
occurs, i.e. $\mu^2$ in eq.(\ref{e2.9a}) is always positive. However, sometimes
eq.(\ref{e2.9b}) forces $|B \mu|$ to be so large that the Higgs potential
becomes unbounded at high scales, i.e. condition (\ref{e2.5}) is violated
at scale \mx. This is not acceptable, since it would lead to a mass of the $Z$
boson of order \mx\ or more.
\item Searches for charginos and neutralinos at LEP, as well as measurements
of the total and invisible width of the $Z$ boson, exclude \cite{31} some
combinations of $M_2, \ \mu$ and \tanb.
\item Higgs searches at LEP \cite{31} exclude scenarios with a light scalar
Higgs boson $h$. As we shall see, $m_P$ is automatically quite large here,
so searches for associate $h P$ production are not relevant. (They do play
a role if \tanb\ is very close to its upper bound, since then $m_P$ can
become very small, as discussed in the previous subsection.)
\item We have required all charged sparticles to be heavier than 45 GeV,
which satisfies all LEP search limits \cite{31}.\footnote{The light stop might
be even lighter if the $\tilde{t}_L-\tilde{t}_R$ mixing angle is chosen such
that the $Z \tilde{t}_1 \tilde{t}_1$ coupling is very small and in addition
\mzi\ is very close to \mst. We have ignored this somewhat artificial
possibility, although solutions of this type can be found in the MSSM
\cite{35}.} Similarly, $m_{\tilde{\nu}} > 40$ GeV makes sure that sneutrinos
contribute little to the invisible width of the $Z$.
\item Searches for squarks and gluinos at hadron colliders \cite{31} are less
straightforward to interpret, since the resulting bounds depend \cite{36} on
the $\tilde{g}$ and $\tilde{q}$ decay branching ratios. We have conservatively
required $\mg \geq 120$ GeV; in most cases the resulting bound on \mhalf\ is
weaker than that from LEP searches.
\item We have required the LSP to be electrically neutral. This is necessary
\cite{37} if the LSP is absolutely stable, which it is in the MSSM as defined
here, since stable charged particles in the interesting mass range would
accumulate in matter, bound to nuclei, leading to an unacceptably large
concentration of exotic isotopes. In practice this is not very constraining,
since \zino\ is almost always the LSP anyway.\footnote{Strictly speaking this
requirement may not be necessary even in our restrictive version of the MSSM,
since the LSP might be the gravitino or some member of the ``hidden sector"
\cite{2} thought to be responsible for SUSY breaking in supergravity or
superstring theories. However, in this case bounds from nucleosynthesis
and from the relic density of the LSP force the LSP mass to be significantly
below the mass of the lightest sparticle in the visible sector, making this
scenario less attractive \cite{38}.}
\item We have required that the scalar potential should not have an absolute
minimum in the $D-$flat directions \cite{39} $\tilde{t}_L = \tilde{t}_R =
H_2^0, \ \tilde{b}_L = \tilde{b}_R = H_1^0$ and $\wt{\tau}_L = \wt{\tau}_R =
H_1^0$ (with all other vevs being zero). This is certainly a necessary
condition if one wants the vacuum we are living in to be absolutely stable. We
are not aware of any readily implementable sufficient condition that would
guarantee the absence of such ``false vacua". This constraint usually limits
$|A|/m_0$ to be below $\sim 3 - 5$.\footnote{Notice that this constraint
should only be imposed at the weak scale. Even if the potential renormalized
at scale \mx\ did have a minimum in one of these directions, the vevs would
still be very roughly of order of the weak scale, so the natural scale to
study them is this lower scale \cite{22}.}
\item For the case $\tanb=8$ we have also required that the LSP relic density
$\Omega h^2$ does not exceed 1, which is roughly equivalent to requiring the
universe to be at least 10 billion years old. Since this cosmological limit
is often considered to be less watertight than the direct search limits and
internal consistency conditions listed above, we show results with (dotted)
and without (solid) this additional constraint. For the purposes of this
article it should be sufficient to keep in mind that the relic density is
approximately inversely proportional to the total LSP annihilation cross
section for non--relativistic LSP's.\footnote{We have computed the LSP
annihilation cross section including all accessible two--body final states
\cite{40}, but used a simple analytical approximation \cite{41} to compute the
relic density from the annihilation cross section. It has recently been
pointed out \cite{42} that a more sophisticated method \cite{43} is necessary
to obtain reliable results in the vicinity of $s-$channel poles in the LSP
annihilation cross section.}
\item Finally, in order to keep the search region finite we have required
$m_0 \leq 1$ TeV and $\mhalf \leq 0.5$ TeV, leading to
$m_{\tilde{g},\tilde{q}} \leq 1.3$ TeV. This is certainly a very crude
``naturalness" criterion, but hopefully not too far from what most practicing
supersymmetrists consider reasonable. In any case, it should be quite
straightforward to extend the allowed regions towards larger masses. We have
indicated the upper ends of these regions in figs.~5, since their dependence
on \tanb\ contains useful information. We will come back to the question of
upper bounds on sparticle masses in sec.~2f.
\end{itemize}

\setcounter{footnote}{0}
Notice that we have not included any constraints from electroweak precision
experiments, or from sparticle loop contributions to rare processes like the
recently popular $b \rightarrow s \gamma$ decays. The reason is that in most
cases sparticle loop contributions are quite small within the MSSM with
radiative symmetry breaking \cite{43a,43b}. Moreover, these calculations
usually contain some theoretical ambiguity. Therefore the region of parameter
space that could be excluded reliably from such considerations is quite small
once the direct search constraints listed above have been incorporated.

Fig.~5a shows the $(M_2,\mu)$ plane often used in experimental as well as
phenomenological studies involving charginos or neutralinos. Notice the
different scales of the two axes. Indeed, probably the most important result
of this figure is that
\be \label{e2.22}
|\mu| \geq 1.4 M_2 \simeq 2.8 M_1
\ee
which holds for our choice $\mt(\mt)=165$ GeV. This implies that \zino\ is a
rather pure gaugino, and even \wino\ and $\wt{Z}_2$ are dominated by their
gaugino components. $\wt{Z}_{3,4}$ and $\wt{W}_2$ are higgsino--like, and
become nearly degenerate in mass as $|\mu|$ increases. This has ramifications
not only for direct searches for all these sparticles, but also for cascade
decays \cite{36} of heavy squarks and gluinos, which usually preferably
produce gaugino--like charginos and neutralinos. It should be noted, however,
that the inequality (\ref{e2.22}) is quite sensitive to \mt; if $\mt(\mt) \leq
155$ GeV, solutions with $|\mu| \leq M_1$, i.e. higgsino--like LSP, can still
be found \cite{40} if \tanb\ is not small and $m_0^2 \gg m^2_{1/2}$.

Notice the overall decrease of $|\mu|$ as \tanb\ is increased from its lower
bound; we have already seen this behaviour in fig.~4b. Finally, requiring a
sufficiently small LSP relic density excludes the region of small or
moderate $M_2$ and very large $|\mu|$. Increasing $|\mu|$ reduces the
coupling of the LSP to $Z$ and Higgs bosons, which usually decreases the LSP
annihilation cross section, i.e. increases the relic density. In addition,
large $|\mu|$ are associated with large $m_0$ here, see eq.(\ref{e2.9a}),
which suppresses LSP annihilation via sfermion exchange in the $t-$channel.

\makebox[15.5cm]{\epsfxsize=0.67\hsize\epsffile{sumo5a.ps}}

\noindent {\bf Fig.~5a:} {\small The allowed region of the $(M_2,\mu)$ plane
allowed in the MSSM for $\tanb = 1.65$ (long dashed), 8 (solid and dotted),
and 35 (short dashed). We have fixed $\mt(\mt)=165$ GeV and varied the SUSY
breaking parameters $m_0, \ \mhalf$ and $A$ within their experimentally and
theoretically allowed ranges, as described in the text. For $\tanb=8$, only
the region enclosed by the dotted curve is allowed by the LSP density
constraint; similar reductions of the allowed parameters space could be
derived from this constraint for our other choices of \tanb. (At small
$|\mu|$, the dotted lines coincide with the solid ones.) Note that the
upper limits in this and the following figures are determined from our rather
arbitrary naturalness condition $m_0 \leq$ 1 TeV, $\mhalf \leq 0.5$ TeV; see
sec.~2f for a detailed discussion of upper limits on sparticle masses.}\\

\makebox[15.5cm]{\epsfxsize=0.67\hsize\epsffile{sumo5b.ps}}

\noindent {\bf Fig.~5b:} {\small The allowed region in the $(\msul,\mst)$
plane. Notations are as in Fig.~5a.}\\

In fig.~5b we show the $(\msul,\mst)$ plane. Here the allowed region obviously
moves only little when \tanb\ is varied. The fixed point solution $\tanb=1.65$
has a slightly reduced maximal value of \mst, due to the increase of the terms
$\propto h_t^2$ in the RGE for $m^2_{\tilde{t}_{L,R}}$; however, this value
obviously depends on our rather arbitrary upper bounds on $m_0$ and \mhalf.
More interesting is the reduction of the allowed region left after the relic
density constraint is imposed. This excludes combinations of parameters where
$m_0$ is large and \mzi\ lies above the range where $s-$channel $H_2$ and $Z$
exchange diagrams are important, unless $\zino \zino \ra \ttbar$ is possible
and has a sizable cross section, which gives an upper bound on \mst\ if
$\msul \geq 1.1 m_0$.\footnote{The increase of the lower bound on \mst\ for
$\msul \geq 600$ GeV is most likely an artifact of our approximate treatment
of $s-$channel poles.}

The most important feature of fig.~5b is that \mst\ can be very much lighter
than \msul; in fact, it can be close to its experimental lower bound even if
$\msul \simeq 1$ TeV. This emphasizes the necessity to take squark
non--degeneracy seriously, e.g. when analyzing squark searches at colliders.

Fig.~5c shows the $(\mg,\mst)$ plane. Here the dependence on \tanb\ is more
pronounced; in particular, the allowed region for $\tanb=1.65$ becomes quite
narrow. The reason is \cite{44} that in this scenario the value of $A_t$ at
the weak scale depends only very weakly on the value of the input parameter
$A$; $A_t$ approaches a fixed point that depends only on \mhalf. For the
other values of \tanb, or if $\mg \leq 600$ GeV, the lower bound on \mst\ is
basically given by $\max(45 \ {\rm GeV}, \mzi)$. For $\mg \geq 450$
GeV, where $s-$channel contributions to \zino\ annihilation become
unimportant, the bound on the LSP relic density reduces the maximal value of
\mst\ by imposing an upper bound on $m_0$, as discussed above. Notice that,
especially for $\tanb=1.6$, $\tilde{g} \rightarrow t \tilde{t}_1$ decays are
often allowed even if the gluino cannot decay into first generation squarks.

\makebox[15.5cm]{\epsfxsize=0.67\hsize\epsffile{sumo5c.ps}}

\noindent {\bf Fig.~5c:} {\small The allowed region in the $(\mg,\mst)$ plane.
Notations are as in Fig.~5a.}\\
\vspace*{4mm}

Fig.~5d depicts the $(\msul,\mg)$ plane. Neither of these masses is affected
by radiative symmetry breaking, so that the allowed region depends very little
on \tanb. The exception is the lower limit on \mg\ (see also the previous
figure): The hadron collider bound $\mg \geq 120$ GeV can be saturated only
for small \tanb\ (and $\mu <0$), since otherwise the chargino search limit
{}from LEP requires $\mg \geq 170$ GeV. Notice that imposing the relic density
constraint greatly reduces the allowed region: If $\mg \geq 450$ GeV (so that
$s-$channel diagrams become unimportant) and $\mg \leq 1$ TeV (so that
$\mzi < \mt$), a rather stringent upper bound on $m_0$ emerges; as a result,
first and second generation squark masses cannot exceed \mg\ significantly.
If $\mg \simeq \msq$, the small mass splittings between squarks
of different flavors, eq.(\ref{e2.16d})--(\ref{e2.16g}), can change
$\tilde{g}$ branching ratios significantly. For example, the larger masses of
$SU(2)$ doublet squarks compared to the $SU(2)$ singlets can greatly reduce
the production of \wino\ and $\wt{Z}_2$ in $\tilde{g}$ decays in such a
scenario. If $\mg > 1$ TeV, \zino\ annihilation into \ttbar\ pair becomes
possible and large values of $m_0$ are allowed as long as \mst\ is sufficiently
small, as discussed above.

\makebox[15.5cm]{\epsfxsize=0.67\hsize\epsffile{sumo5d.ps}}

\noindent {\bf Fig.~5d:} {\small The allowed region of the $(\msul,\mg)$ plane.
Notations are as in Fig.~5a.}\\
\vspace*{4mm}

We next turn to the $(m_P,\msul)$ plane shown in fig.~5e. For fixed \tanb\ the
correlation between these masses is quite striking; it is a direct consequence
of radiative gauge symmetry breaking. This correlation becomes even stronger
if we impose the relic density constraint. However, the strong dependence of
$m_P$ on \tanb, which we have already seen in fig.~4b, means that the
correlation shown in fig.~5e can only be used to test the MSSM if \tanb\ is
known. Notice finally that even for $\tanb=1.65$ the pseudoscalar Higgs boson
is never heavy enough to decay into a pair of first or second generation
squarks.

In contrast, fig.~5f shows that scenarios with $m_P > 2 \mst$ can quite
easily be found. While the pseudoscalar Higgs boson does not couple to an
identical squark--antisquark pair, the heavy scalar Higgs boson, whose mass
is very close to $m_P$, does \cite{16}. One can even find cases where
$m_P > \mst + m_{\tilde{t}_2}$; in this case $H^+ \ra \sbo \tilde{t}_1$ will
also occur. In some cases these decays into stops might even constitute the
dominant decay modes for the heavy Higgs bosons of the MSSM. Finally, we again
observe the reduction of $m_P$ with increasing \tanb, and the shrinking of
the allowed parameter space with large \mst\ if the LSP relic density
constraint is imposed.

\makebox[15.5cm]{\epsfxsize=0.67\hsize\epsffile{sumo5e.ps}}

\noindent {\bf Fig.~5e} {\small The allowed region in the $(m_P,\msul)$ plane.
Notations are as in Fig.~5a.}\\

\makebox[15.5cm]{\epsfxsize=0.67\hsize\epsffile{sumo5f.ps}}

\noindent {\bf Fig.~5f} {\small The allowed region in the $(m_P,\mst)$ plane.
Notations are as in Fig.~5a.}\\
\vspace*{4mm}

Fig.~5g shows that there exists only a mild correlation between $m_P$ and the
LSP mass \mzi, unless the relic density constrained is imposed, which greatly
reduces the upper limit of $m_P$ for $\mzi < \mt$. Note that even for
$\tanb=35$ we always have $m_P > 2 \mzi$. This means that $P-$exchange
contributions to the LSP annihilation cross section will still be suppressed.
More importantly, it also means that $P$ (and hence also the heavy scalar
$H_1$) will have some SUSY decay modes, unless $\tanb \geq 40$ or so. This
casts doubt on the reliability of studies of SUSY Higgs searches that ignore
Higgs decays into sparticles.

\makebox[15.5cm]{\epsfxsize=0.635\hsize\epsffile{sumo5g.ps}}

\noindent {\bf Fig.~5g:} {\small The allowed region in the $(m_P,\mzi)$ plane.
Notations are as in Fig.~5a.}\\
\vspace*{4mm}

Fig.~5h depicts the $(\mst,\mzi)$ plane. Since \zino\ is always gaugino--like,
this figure is essentially a re--scaled (and rotated) version of fig.~5c. We
include it mostly in order to emphasize that even in the MSSM, the
$\tilde{t}_1 - \zino$ mass difference might be very small. The open production
of pairs of light stop squarks that are nearly degenerate with the LSP might
be very difficult to detect, especially at hadron colliders. This may be an
example where even a strongly interacting particle is more easily detected
at $e^+e^-$ colliders. (However, $\tilde{t}_1 \tilde{t}_1^*$ bound states
might be detectable \cite{45} at hadron colliders in such a scenario.) Note
also that in some cases $t \rightarrow \tilde{t}_1 \zino$ decays are possible.

\makebox[15.5cm]{\epsfxsize=0.635\hsize\epsffile{sumo5h.ps}}

\noindent {\bf Fig.~5h:} {\small The allowed region in the $(\mst,\mzi)$ plane.
Notations are as in Fig.~5a.}\\

Fig.~5i shows that there is little correlation between \mstau\ and \mwi\
unless we impose the relic density constraint. As discussed above, this leads
to a strong upper bound on $m_0$ if 55 GeV $\leq \mzi \leq \mt$, which
roughly translates to 110 GeV $\leq \mwi \leq 2 \mt$ since $\mzi \simeq M_1$
and $\mwi \simeq M_2$ are related by eq.(\ref{e2.18}). Similarly, the
requirement $\mstau > \mzi$ defines the upper bound on \mwi\ for given
\mstau. Notice that even if the relic density constraint is imposed, we can
find examples both where $\wino \ra \stau + \nu_{\tau}$ decays are allowed and
those where $\stau \ra \wino + \nu_{\tau}$ decays are possible. Finally, the
reduction of \mstau\ due to terms involving $h_{\tau}$ is still quite small
even for $\tanb=35$.

\makebox[15.5cm]{\epsfxsize=0.65\hsize\epsffile{sumo5i.ps}}

\noindent {\bf Fig.~5i:} {\small The allowed region in the $(\mstau,\mwi)$
plane. Notations are as in Fig.~5a.}\\
\vspace*{4mm}

\makebox[15.5cm]{\epsfxsize=0.65\hsize\epsffile{sumo5j.ps}}

\noindent {\bf Fig.~5j} {\small The allowed region in the $(\msul,\mstau)$
plane. Notations are as in Fig.~5a.}\\

This can also be seen from fig.~5j, which shows the $(\msul,\mstau)$ plane.
Nearly the entire plane with $1 \leq \msul/\mstau \leq 5$ is allowed unless
the relic density constraint is imposed. Notice that the lower bound on
\mstau\ for given \msul\ is set by the requirement $\mstau > \mzi$,
independent of \tanb; on the other hand, the upper limit on \mstau\ for fixed
\msul\ is reduced somewhat for large \tanb, as already shown in fig.~4b.

This concludes our pictorial exploration of the spectrum of the MSSM. We
already mentioned that we cannot derive upper bounds on (s)particle masses
{}from these figures, since we imposed arbitrary upper limits on SUSY breaking
mass parameters. However, at least for fixed \tanb\ one can often give lower
bounds on (s)particle masses that are significantly more stringent than the
direct search limit for this (s)particle. This simply results from the fact
that the model contains far fewer free parameters than new (s)particles; any
experimental lower bound can therefore have ramifications for the allowed
range of other masses.

This is demonstrated in table~1, where we give the lower bounds on $|\mu|$
and $M_2$ as well as several sparticle and Higgs boson masses, for the three
values of \tanb\ considered in figs.~5. The only experimental lower bounds
that can be saturated for all values of \tanb\ are those on \mst\ and on
\mwi. On the other hand, LEP Higgs searches are found to constrain the
parameter space only at the lowest value of \tanb\ considered; as remarked
above, they again become relevant at very large \tanb. Table~1 shows that
even the discovery of a Higgs boson at LEP2 would likely exclude a large
range of intermediate values of \tanb, since here the mass of lightest scalar
Higgs boson must exceed 94 GeV in the MSSM; this is related to the lower
bound on $m_P$, which in turn is determined indirectly by the neutralino and
chargino searches at LEP1. On the other hand, the Higgs search constraint from
LEP1 helps to push the minimal possible value of the sneutrino mass to 55 GeV
if $\tanb=1.65$; at this small value of \tanb\ sizable radiative corrections
to $m_{H_2}$ are needed, which puts constraints on the allowed choices of SUSY
breaking parameters.

\vspace*{4mm}
\noindent
{\bf Table~1:} {\small The minimal values of the parameters $M_2$ and $\mu$ and
of various sparticles masses allowed in the MSSM, given the experimental
and theoretical constraints discussed in sec.~2c, but without constraint on
the \zino\ relic density; in case of the light scalar Higgs boson $h$ both
lower and upper bounds are given. The values of \tanb\ chosen here are the
same as in figs.~5. All masses are in GeV.}

\begin{center}
\begin{tabular}{|c||c|c|c|}
\hline
  & $\tanb=1.65$ & $\tanb=8$ & $\tanb=35$ \\
\hline
$|\mu|$ & 198 & 108 & 120 \\
$M_2$   &  32 &  44 &  47 \\
$m_P$   & 241 & 119 &  86 \\
$m_h$   & 60/107 & 94/132 & 85/132 \\
$m_{\tilde{\nu}}$& 55 & 40 & 40  \\
\mstau\ &  49 &  56 &  47 \\
$m_{\tilde{u}_L}$ & 235 & 172 & 221 \\
\msb\   & 216 & 182 & 119 \\
\mst\   &  45 &  45 &  45 \\
\mg\    & 120 & 170 & 181 \\
\mzi\   &  18 &  23 &  24 \\
\mwi\   &  45 &  45 &  45 \\
\hline
\end{tabular}
\end{center}

Conversely, if the production of first or second generation squarks is
observed at the Tevatron in the near future, very large or very small values
of \tanb\ could be excluded within the MSSM, since here the lower bound on
\msul\ is quite high. (Bounds for other first or second generation squark
masses are very similar.) The bound on \msul\ is always at least partly
determined by chargino and neutralino searches at LEP. In addition, at low
\tanb\ Higgs searches are important for determining the minimal allowed value
of \msul, while at high \tanb\ the \stau\ search bound from LEP becomes
relevant. On the other hand, while the present bound on the gluino mass can
only be saturated at small \tanb, with increased luminosity the Tevatron
should be able to probe some part of the allowed parameter space via gluino
searches even for larger values of \tanb. Notice that the \sbo\ can also be
quite light if \tanb\ is large.

The rather large lower bounds on $|\mu|$ imply that within the MSSM one would
not expect the heavier chargino or either of the heavier neutralinos to be
detectable at LEP2. On the other hand, searches for the pair production of
light charginos, for associate $\zino \wt{Z}_2$ production, and for the
production of slepton or stop pairs might yet succeed, or at least would allow
to derive new constraints of the parameter space of the MSSM. We thus see that
even in this rather constrained model it is quite possible that existing
facilities (LEP, Tevatron) will discover a SUSY signal before the LHC is
completed. On the other hand, a glance at figs.~5 shows that a discovery
cannot be guaranteed at these colliders.

So far our explanations for the features observed in figs.~4 and 5 as well as
in table~1 have been rather qualitative. In the next subsection we will discuss
analytical approximation that allow a semi--quantitative understanding of the
MSSM with the help of a pocket calculator. Readers with access to a program
that computes the sparticle spectrum, or who are only interested in numerical
predictions of the MSSM, might want to skip this subsection and directly
continue with our discussion of some sample spectra in sec.~2e.

\setcounter{footnote}{0}
\subsection*{2d) Approximate Analytical Expressions}
In this subsection we give approximate analytical solutions of the the
relevant RGE. Together with eq.(\ref{e2.9a}), which determines the value of
$|\mu|$, they allow one to compute sparticle and Higgs masses to a precision
of typically 10 to 20\%.

We have already seen that the RGE for running scalar masses are easily
integrated if all Yukawa couplings are negligible compared to the gauge
couplings; the results are listed in eqs.(\ref{e2.16}). If there is only one
sizable Yukawa coupling ($h_t$), the solutions for the running Higgs and
squark masses can still be expressed \cite{28} in terms of a single definite
integral that has to be computed numerically. No such solutions can be given
if two or more Yukawa couplings have to be taken into account. However, in
practice the solutions of the RGE can still be approximated by sums of terms
proportional to a single squared Yukawa coupling; the cross--interactions
between these terms is quite small, partly due to the smallness of certain
numerical coefficients in the RGE \cite{15}.\footnote{For example, the r.h.s.
of the RGE for $h_t^2$ contains the combination $6 h_t^2 + h_b^2$, while in
the RGE for $h_b^2$ one finds the term $6 h_b^2 + h_t^2$; the Yukawa couplings
affect each other's running therefore only rather weakly.}

We therefore start with the ansatz:
\ben \label{e2.23} \beq
m^2_{H_1} &= m_0^2 + 0.52 m^2_{1/2} - X_b - X_{\tau}/3 ; \label{e2.23a} \\
m^2_{H_2} &= m_0^2 + 0.52 m^2_{1/2} - X_t . \label{e2.23b}
\eeq \een
In the (unrealistic) limit of vanishing Yukawa couplings, $X_t=X_b=X_{\tau} =
0$, and the SUSY breaking Higgs mass parameters run just like sneutrino
masses. The same quantities $X_i$ also appear in the running third generation
sfermion masses:
\ben \label{e2.24} \beq
m^2_{\tilde{\tau}_R} &= m_0^2 + 0.15 m^2_{1/2} - \frac{2}{3} X_{\tau};
\label{e2.24a} \\
m^2_{\tilde{\nu}_{\tau}} = m^2_{\tilde{\tau}_L} &= m_0^2 + 0.52 m^2_{1/2}
- \frac{1}{3} X_{\tau}; \label{e2.24b} \\
m^2_{\tilde{t}_L} = m^2_{\tilde{b}_L} &= m_0^2 + (0.47 + \cg) m^2_{1/2}
-\frac{1}{3} ( X_b + X_t); \label{e2.24c} \\
m^2_{\tilde{t}_R} &= m_0^2 + (0.07 + \cg) m^2_{1/2} -\frac{2}{3} X_t;
\label{e2.24d} \\
m^2_{\tilde{b}_R} &= m_0^2 + (0.02 + \cg) m^2_{1/2} -\frac{2}{3} X_b,
\label{e2.24e}
\eeq \een
where \cg\ has been given in eq.(\ref{e2.17}). Approximate expressions for the
$X_i$ are:\footnote{A similar expression for $X_t$ has been found by Carena et
al. \cite{44}; for approximate expressions including effects of the bottom
Yukawa coupling, see ref.\cite{45a}.}
\ben \label{e2.25} \beq
X_t &= \left( \frac {\mt(\mt)} {150 \ {\rm GeV} \sin \! \beta} \right)^2
\left\{ \rule{0mm}{8mm} 0.9 m_0^2 + 2.1 m^2_{1/2} \non \right. \\
& \left. \hspace*{4cm} + \left[ 1 - \left( \frac {\mt(\mt)} {190 \ {\rm GeV}
\sin \! \beta} \right)^3 \right] \left( 0.24 A^2 + A \mhalf \right) \right\};
\label{e2.25a} \\
X_b&= X_t (\mt(\mt) \leftrightarrow m_b(\mt), \ \sin \! \beta \leftrightarrow
\cos \! \beta) ; \label{e2.25b} \\
X_{\tau} &= \frac {10^{-4}} {\cos^2 \beta} \left( m_0^2 + 0.15 m^2_{1/2}
+ 0.33 A^2 \right). \label{e2.25c}
\eeq \een
The only other quantity needed to compute the sparticle spectrum is
\be \label{e2.26}
A_t = A \left[ 1 - \left( \frac {\mt(\mt)} {190 \ {\rm GeV} \sin \! \beta}
 \right)^2 \right] + \mhalf \left[ 3.47 - 1.9 \left( \frac {\mt(\mt)} {190 \
{\rm GeV} \sin \! \beta} \right)^2 \right].
\ee
Whenever $\tilde{b}_L-\tilde{b}_R$ or $\wt{\tau}_L-\wt{\tau}_R$ mixing are
important they are dominated by the term $\propto \mu$ in eq.(\ref{e2.12b}),
since it grows $\propto \tanb$.

Notice that $|\mu|$ practically only depends on $X_t$; the terms $\propto
X_{b,\tau}$ in eq.(\ref{e2.9a}) come with a factor $\cos^2 \beta$ and are
thus always negligible. Inserting eqs.(\ref{e2.23}) and (\ref{e2.25a}) into
eq.(\ref{e2.9a}) gives:
\be \label{e2.27}
\mu^2 \simeq -m_0^2 - 0.52 m^2_{1/2} - \frac{1}{2} m_Z^2 + \frac {X_t}
{1 - \cot^2 \beta},
\ee
which holds at the weak scale. For $\mt(\mt) > 158$ GeV the coefficient of
$m_0^2$ on the r.h.s. of eq.(\ref{e2.27}) is always positive; the coefficient
of $m^2_{1/2}$ is positive if $\mt>75$ GeV. This explains why $|\mu|$ is
always quite large for our standard choice $\mt(\mt)=165$ GeV, but also
indicates that arbitrarily small values of $|\mu|$ will be allowed for small
ratio $\mhalf/m_0$ if $\mt(\mt) < 155$ GeV, which is less than $1\sigma$ below
the central value cited by the CDF collaboration \cite{17}.

Another useful observation \cite{25} is that the mass of the pseudoscalar
Higgs boson, eq.(\ref{e2.11}), is independent of $X_t$ if it is expressed in
terms of $\mu^2$:
\be \label{e2.28}
m_P^2 \simeq \frac { m^2_{\tilde{\nu}} + \mu^2 - X_b - X_{\tau}/3}
{\sin^2 \beta}.
\ee
For $\tanb \leq 20$ or so, $X_b$ and $X_{\tau}$ are essentially negligible;
the resulting simple expression is useful since it establishes a quantitative
connection\footnote{Corrections stemming from the 1--loop potential
(\ref{e2.15}) only amount to a few percent if the scale $Q_0$ is chosen to be
of the order of the average stop mass, as discussed at the end of sec.~2a.}
between a sparticle mass, a Higgs mass and the $\mu$ parameter. This is to be
contrasted with many phenomenological analyses where the parameters of the
sfermion, chargino/neutralino, and Higgs sectors are often considered to be
completely independent. Eq.(\ref{e2.28}) means that in the MSSM the
pseudoscalar Higgs boson, and hence also the heavy neutral scalar as well as
the charged Higgs boson, are heavier than the higgsino--like chargino
and neutralinos, whose masses are close to $|\mu|$, and also heavier than the
sleptons, {\em unless} \tanb\ is very large. This explains our observation of
the previous subsection that for $\tanb \leq 35$, the heavy neutral Higgs
bosons always have some supersymmetric decay channels.

\subsection*{2e) Sample Spectra}
In this subsection we present a few examples of (s)particle spectra as
computed from the four input parameters of the MSSM ($m_0, \ \mhalf, \ A$
and \tanb). In the previous subsection we emphasized that the mechanism of
radiative symmetry breaking allows one to compute the value of $|\mu|$
{}from this input; together with the boundary condition (\ref{e2.7}) this
also determines the Higgs boson masses. The MSSM as defined here has
therefore fewer free parameter than has often been assumed in
phenomenological analyses, where $\mu$ and $m_P$ have been varied
independently of the sfermion and gaugino masses.

In this subsection we want to emphasize that in spite of this small
number of free parameters the MSSM allows to construct more complicated
spectra at the weak scale than have usually been considered. This is
partly due to the possible reduction of third generation (current) sfermion
masses by the Yukawa terms in the RGE, and partly due to mixing between
$SU(2)$ doublet and singlet sfermions of the same charge; the first effect
lowers the average mass of the two sfermion eigenstates with a given
flavor, while the second effect leads to a mass splitting between these
two eigenstates. The overall result is that some third generation sfermions
can be significantly lighter than their counterparts in the first two
generations. Since third generation sfermions almost always decay into
third generation fermions, which lead to quite different signatures than
first or second generation fermions, such mass splittings can have quite
dramatic consequences for the kind of SUSY signals one expects at a
given collider.

Both the reduction of sfermion current masses and mixing between
current eigenstates becomes more important with increasing Yukawa
couplings. By now we know that the top quark is heavy, so that its
Yukawa coupling is ${\cal O}(1)$. Both effects are therefore usually
important for stop squarks.\footnote{$\tilde{t}_L-\tilde{t}_R$ mixing
can be ``accidentally" small if $A_t + \mu \cot \! \beta$ is small due
to a cancellation, see eq.(\ref{e2.12a}).} The possibility that the
lighter stop eigenstate lies significantly below the other squarks has
already been discussed in a number of papers \cite{46,47}. However, it
has not always been appreciated that the top Yukawa coupling also
reduces the mass of the $SU(2)$ partner of $\tilde{t}_L$, the
$\tilde{b}_L$, compared to those of $\tilde{d}_L$ and $\tilde{s}_L$;
this is a direct consequence of $SU(2)$ invariance. If \tanb\ is not
large, $h_b$ and $h_{\tau}$ are quite small; in this case there is
little mixing between the superpartners of left-- and right--handed
$b$'s and $\tau$'s, and $\tilde{b}_2$ and $\wt{\tau}_{1,2}$ will be
close in mass to the corresponding first generation sfermions.

An example for such a ``typical" scenario is given in the first
column of table~2, where we have taken $m_0 = \mhalf = 100$ GeV,
$A=0, \ \tanb=2$ and $\mu<0$. (Recall that radiative symmetry
breaking only fixes $|\mu|$, but leaves its sign free.) This leads
to first generation squark and gluino masses at the upper end of
the range that might eventually be detectable at the Tevatron. Since
the SUSY breaking scale is rather low, the supersymmetric contributions
to the stop mass matrix (\ref{e2.12a}), i.e. the $m_t^2$ terms on the
diagonal, are not negligible; they largely cancel the reduction of the
diagonal SUSY breaking mass terms due to the top Yukawa coupling.
Moreover, since $\mu<0$ and \tanb\ is quite small there is a partial
cancellation in the off--diagonal terms; note that even though $A=0$
at the GUT scale, $A_t>0$ at the weak scale due to gaugino loops
\cite{15}. As a result, $\tilde{t}_1$ is not much lighter than first generation
squarks here, while $\tilde{t}_2$ is the heaviest of all squarks.

The lighter $\tilde{b}$ eigenstate is (coincidentally) very close to
$\tilde{t}_1$ in mass in this example. This rather moderate decrease of
\msb\ compared to first generation squark masses is quite important, however,
since it increases the branching ratio for $\tilde{g} \ra \tilde{b} b$
decays to about 35\%. Since $\tilde{q} \tilde{g}$ and $\tilde{g} \tilde{g}$
cross sections are larger than squark pair cross sections at hadron colliders,
this means that in this scenario SUSY events at hadron colliders are
expected to be $b-$rich. Moreover, since \sbo\ is mostly an $SU(2)$ doublet,
the reduction of \msb\ leads to an enhanced rate for $\wt{Z}_2$ production
in \gl\ decays, compared to the case where all squark masses are equal;
note that $\sbo \ra t \wt{W}_1$ is kinematically forbidden here.

In this example sleptons are substantially lighter than squarks; this is always
true in the MSSM, unless $m_0^2 \gg m^2_{1/2}$. Because of the rather low SUSY
breaking scale the $D-$term contributions to eqs.(\ref{e2.16a})--(\ref{e2.16c})
are still of some importance, leading to some $\tilde{e}_L - \tilde{\nu}$ mass
splitting; the near--degeneracy of $\tilde{e}_R$ and $\tilde{\nu}$ in this
example is accidental. For larger SUSY breaking scales $\tilde{e}_R$ will be
the lightest first generation slepton, again unless $m_0^2 \gg m^2_{1/2}$ in
which case all first generation sleptons are very close in mass. Since
$h_{\tau}$ is small, the $\wt{\tau}$ eigenstates lie very close to the
$\tilde{e}$ and $\tilde{\mu}$ here.

Because we have chosen $\mu < 0$ and small \tanb, higgsino--gaugino mixing
increases \mwi\ and \mzit\ above the current mass $M_2 \simeq 84$ GeV;
\mzi\ is also somewhat larger than $M_1 \simeq 42$ GeV. As a result, the
only SUSY process that might be detectable at LEP2 (with $\sqrt{s} \leq 200$
GeV) is $\wt{Z}_1 \wt{Z}_2$ production; since the selectrons are quite
light, the cross section for this process should not be too small. Finally, the
light scalar Higgs $h$ should be easily detectable at LEP2 in this scenario.

In the second column of table~2 we have chosen the same values for the
dimensionful parameters $m_0, \ \mhalf$ and $A$, but we have increased
\tanb\ to 40. This has quite dramatic effects on the resulting
phenomenology; in particular, even though all dimensionful input parameters
(and hence ``the SUSY breaking scale") have remained the same, many
sparticles have become significantly lighter.\footnote{Unfortunately the
ISASUSY program package that computes branching ratios for various sparticle
and Higgs decay modes in the MSSM does not yet allow for mass splitting
between different generations of sleptons. The following discussion therefore
has to remain qualitative at times.}

\vspace*{4mm}
\noindent
{\bf Table~2:} {\small Three MSSM sample spectra for $\mt(\mt)=165$ GeV; the
values of the other input parameters are as indicated. $\tilde{u}_R$ and
$\tilde{d}_L$ are the lightest and heaviest first generation squark,
respectively. All masses are in GeV.}
\begin{center}
\begin{tabular}{|c||c|c|c|}
\hline
 & Case 1 & Case 2 & Case 3 \\
\hline
$m_0$       & 100 & 100 & 200 \\
\mhalf\     & 100 & 100 & 100 \\
$A$         &   0 &   0 & 400 \\
\tanb\      &   2 &  40 &  40 \\
$\mu$       & -200 & -134 & -211 \\
\hline
$m_P$       & 256 &  76 &  99 \\
$m_h$       &  66 &  75 &  95 \\
$m_H$       & 268 & 104 & 110 \\
$m_{H^+}$   & 268 & 110 & 127 \\
\hline
\mzi\       &  46 &  37 &  41 \\
\mzit\      &  97 &  64 &  74 \\
$m_{\tilde{Z}_3}$ & 207 & 154 & 226 \\
$m_{\tilde{Z}_4}$ & 225 & 178 & 237 \\
\mwi\       &  97 &  62 &  74 \\
$m_{\tilde{W}_2}$ & 224 & 183 & 242 \\
\hline
$m_{\tilde{e}_R}$ & 113 & 116 & 208 \\
$m_{\tilde{e}_L}$ & 128 & 132 & 217 \\
$m_{\tilde{\nu}}$ & 112 & 105 & 202 \\
\mstau\     & 112 &  63 & 120 \\
$m_{\tilde{\tau}_2}$ & 128 & 149 & 211 \\
\hline
\mg\        & 272 & 271 & 275 \\
$m_{\tilde{u}_R}$ & 247 & 246 & 299 \\
$m_{\tilde{d}_L}$ & 261 & 263 & 313 \\
\msb\       & 233 & 186 & 156 \\
$m_{\tilde{b}_2}$ & 250 & 255 & 267 \\
\mst\       & 231 & 190 & 126 \\
$m_{\tilde{t}_2}$ & 287 & 313 & 325 \\
\hline
\end{tabular}
\end{center}
\vspace*{4mm}

To begin with, \mst\ has gone down, mostly since there is little cancellation
in the off--diagonal entry of the mass matrix (\ref{e2.12a}) now, the term
$\propto \mu$ being suppressed by a factor $\cot \! \beta = 1/40$; for the
same reason \mstt\ has gone up slightly compared to the previous case. Since
$h_b$ is now quite substantial, the diagonal entries of the sbottom mass
matrix are reduced, while the off--diagonal entries are much larger than
before;
these two effects tend to cancel for the heavier eigenstate, but go in the
same direction for the lighter one. As a result, \sbo\ is the lightest
squark and $Br(\gl \ra \tilde{b} b) \simeq 64\%$.

The masses of first and second generation squarks and sleptons are little
affected by the increase of \tanb, but there is now a very substantial mass
splitting between the lighter $\wt{\tau}$ eigenstate and the other sleptons.
As a result, \stau\ pair production should be easily observable at LEP2 in
this example. Moreover, $\wt{W}_1$ and $\wt{Z}_2$ decays almost always lead
to final states containing $\tau \nu_{\tau}$ and $\tau^+ \tau^-$ pairs,
respectively. This greatly reduces the cross section for the production of
hard electrons or muons in SUSY events at hadron colliders, but should
increase the missing $E_T$ signal due to the large number of $\tau$ neutrinos
in the event; obviously, the capability to detect $\tau$ leptons at a hadron
collider would be very helpful here. Notice also that $\wt{W}_1$ and
$\wt{Z}_2$ are now light enough to be pair produced at LEP2; since $\wt{W}_1$
and \stau\ are very close in mass here, it might prove challenging to
disentangle contributions from $\wt{\tau}_1^+ \wt{\tau}_1^-$ and
$\wt{W}_1^+ \wt{W}_1^-$ pairs to the production of acollinear $\tau^+ \tau^-$
pairs.

The increase of \tanb\ has also decreased $|\mu|$, see eq.(\ref{e2.9a}) or
(\ref{e2.27}), which reduces $m_{\tilde{Z}_3}, \ m_{\tilde{Z}_4}$, and
$m_{\tilde{W}_2}$; even $\wt{Z}_1 \wt{Z}_3$ production might now be detectable
at LEP2. Moreover, the mass of the pseudoscalar Higgs boson has been greatly
reduced compared to the previous example, so that associate $hP$ production
should be easily observable at LEP2. On the other hand, the near--degeneracy
of $h$ and $P$ means that the $ZZh$ coupling is very small \cite{16}; $Zh$
production will therefore probably not be detectable.

The reduced Higgs boson masses also imply greatly increased Higgs production
rates at hadron colliders. For example, the charged Higgs boson is now light
enough to be produced in top decays. Moreover, {\em all} Higgs bosons of the
MSSM can be produced in decays of the heavier stop eigenstate \stot. Even
though \stot\ pairs can be produced via strong interactions and at least some
of the branching ratios of \stot\ into a Higgs boson plus \sto\ or \sbo\
should be sizable, such events might be quite difficult to detect at hadron
colliders; recall that \gl\ decays also lead to $b-$rich events in this
scenario, so that $b-$tagging may not be sufficient to isolate the Higgs
signal. The study of \stot\ decays into Higgs bosons is very interesting
since it probes directly the trilinear scalar interactions, which are not
easily accessible experimentally.

Other sources of Higgs bosons at hadron colliders in this example are the
decays of $\wt{Z}_3$ and $\wt{Z}_4$ produced in \sbo\ decays; the
$Br(\sbo \ra \wt{Z}_{3,4})$ should be quite substantial here, since
$\wt{Z}_{3,4}$ are higgsino--like and $h_b$ is larger than the electroweak
gauge couplings. Both in \stot\ and in $\wt{Z}_{3,4}$ decays modes
containing a Higgs boson have to compete with modes containing a real
$W$ or $Z$ boson.

In the last column of table~2 we have kept \mhalf\ and \tanb\ as in the
second column, but increased $m_0$ and $A$ to 200 and 400 GeV, respectively.
Because of the larger value of $m_0$ all first and second generation squarks
are now heavier than the gluino, and will thus predominantly decay into a
gluino and a light quark. Due to the increase of $A$, \mst\ has gone
down compared to the previous example; even though \sto\ is now again the
lightest squark, $\gl \ra \sto t$ decays are still kinematically forbidden.
As a result, the only two--body decays of \gl\ now involve sbottom squarks,
which means that almost all \gl\ decays will produce at least one $b \bar{b}$
pair.

The increase of $m_0$ and $A$ (with $\mhalf A > 0$) has led to an increase of
$|\mu|$, and thus of the masses of the heavier neutralino and chargino states;
in particular, $\wt{Z}_{3,4}$ are no longer accessible in \sbo\ decays, and
will thus be produced only very rarely in the decays of gluinos or the
superpartners of light quarks. The larger value of $|\mu|$ also reduced
gaugino--higgsino mixing, which slightly increased \mzi, \mzit\ and \mwi\
compared to the previous example; however, $\wt{Z}_2$ and $\wt{W}_1$ can still
be pair--produced at LEP2.

All slepton masses are significantly larger than in the previous example;
in particular, $\wt{W}_1$ and $\wt{Z}_2$ decays can no longer proceed via the
exchange of an almost real \stau. However, since $\wt{Z}_2$ and, to a lesser
extent, $\wt{W}_1$ decays are still dominated by sfermion exchange
contributions and \stau\ is still by far the lightest slepton, the
branching ratio for final states containing $\tau$ leptons is still enhanced
significantly compared to those for electrons or muons; notice that the
branching ratios scale like the inverse fourth power of the mass of the
exchanged sfermion if sfermion exchange contributions dominate.

Finally, the masses of the Higgs bosons have also gone up compared to the
example in column 2; no Higgs signal will be observable at LEP2 in this
scenario unless the center--of--mass energy can be increased to at least
200 GeV. However, $t \ra H^+ b$ decays are still allowed, and all four
Higgs bosons of the MSSM can again be produced in \stot\ decays.

In this subsection we have emphasized examples with large \tanb, since such
scenarios have not yet been explored very much in the literature. We picked
relatively low values of $m_0$ and \mhalf, leading to rather light sparticles,
in order to emphasize that these considerations are of interest to present and
near--future experiments; unfortunately, much larger sparticle masses cannot
be excluded, as will be discussed in the next subsection. Spectra for the case
where the top Yukawa coupling is close to its upper bound (fixed--point
scenario) have been discussed in several recent papers \cite{44,47,48}. Other
``generic" spectra can be found in refs.\cite{29}.

\setcounter{footnote}{0}
\subsection*{2f) An Upper Bound on Sparticle Masses?}
We have emphasized repeatedly that the small number of free parameters makes
the MSSM quite predictive. However, so far we have not addressed in a
quantitative way the perhaps most important question about phenomenological
SUSY: At what mass should we expect sparticles to appear? We have seen in the
examples of the previous subsection, as well as the figures of sec.~2c, that it
is quite possible that some sparticles will be discovered at LEP2 or the
Tevatron; however, such a discovery is by no means guaranteed. Given the (in
our view) strong motivation for SUSY (see sec.~1), it is obviously important to
determine just how high in energy one has to go in order to test this idea
decisively, or at least to test the MSSM.

Actually, we have already mentioned a couple of times that the MSSM predicts
the existence of at least one new particle with relatively moderate mass:
The light scalar Higgs boson $h$ should lie below 135--140 GeV, even if we
interpret bounds on \mt\ conservatively. Given this bound, it can be shown
\cite{49} that an $e^+ e^-$ collider with $\sqrt{s} \geq 300$ GeV would
have to detect at least one of the Higgs bosons of the MSSM; this remains
true even if we do not require radiative gauge symmetry breaking, and allow
for arbitrary boundary conditions at the GUT scale. Such a collider has
therefore the potential to exclude all SUSY models with minimal Higgs
content. We will see in sec.~4b that this no--lose theorem can be
extended to {\em all} SUSY models, as long as the Higgs sector remains
weakly interacting up to scales of order \mx.

Unfortunately the detection of $h$ would not prove the existence of SUSY,
however. If the pseudoscalar Higgs is heavy, which will be true in the
MSSM with heavy sparticles unless \tanb\ is close to its upper bound, $h$
will behave just like the Higgs boson of the nonsupersymmetric SM. Indeed,
{\em any} model where the Higgs sector remains perturbative up to scales of
order \mx\ has at least one neutral scalar Higgs boson with mass below about
200 GeV; if this is the only light Higgs boson, its couplings will always
resemble those of the SM. The upper bound for weakly coupled SUSY theories is
somewhat below that for nonsupersymmetric ones; hence the introduction of
weak--scale SUSY does lead to a definite sharpening of this bound.
Nevertheless, by looking for a light Higgs boson one really tests whether
physics below the Planck scale can be described by a perturbative quantum
field theory, rather than the existence of weak--scale SUSY.\footnote{One
might argue that a perturbative Higgs sector really only makes sense in the
supersymmetric version, where corrections to the Higgs masses are under
control. However, such naturalness arguments are not likely to convince all
skeptics.} There is no substitute for the detection of at least one sparticle.

Three methods have been suggested in the literature to derive (approximate)
upper bounds on sparticle masses: Threshold effects in the running gauge
couplings; naturalness arguments; and bounds on the cosmological relic
density of LSPs left over from the very early universe. We will briefly
discuss all three methods here.

Determining sparticle masses from the running gauge couplings has several
weaknesses. To begin with, one has to assume gauge coupling unification.
This includes some assumption about threshold corrections at the GUT scale;
after all, it makes little sense to include threshold corrections at only
one of the two scales between which the gauge couplings are run. This means
that any bounds on sparticle masses will depend quite strongly on details of
the GUT model one is considering. Finally, sparticle masses enter the running
of the gauge couplings only logarithmically. Allowing for experimental and
theoretical uncertainties, e.g. due to GUT--scale threshold effects, then leads
to a very large uncertainty in the upper bounds on sparticle masses that can be
derived in this fashion. For this reason most recent analyses (see e.g.
\cite{7}) assume a range of sparticle masses, usually from naturalness
arguments, and treat the corresponding range of threshold corrections as
uncertainties in the SUSY GUT predictions for the gauge couplings measured at
scale \mz.

The use of naturalness arguments to derive quantitative upper bounds on
sparticle masses has been pioneered by Barbieri and Giudice \cite{50}. It
should be noted that neither ref.\cite{50} nor later refinements of this
method \cite{51} use radiative corrections to Higgs masses as a naturalness
criterion, which would follow most closely the main motivation for weak--scale
SUSY. Rather, these analyses always assume radiative gauge symmetry breaking,
and require that the spectrum at the weak scale, including the $Z$ boson
mass, should not depend too sensitively on the values of the input parameters.
The most sensitive parameter is usually the top quark Yukawa coupling $h_t$.

The main advantage of this method is that sparticle masses enter the
fine--tuning constraint quadratically, as opposed to the logarithmic
dependence of the threshold corrections to the running gauge couplings. The
reason is that the Higgs vevs, and hence \mz, are determined essentially by
the dimensionful parameters of the Higgs potential (\ref{e2.4}), all of which
depend quadratically on some SUSY breaking parameter (or $\mu$). The most
``natural" values of the vevs would then be of the order of those masses (or
zero or infinite, if conditions (\ref{e2.5}) or (\ref{e2.6}) are violated). If
these masses are significantly larger than \mz, $h_t(\mx)$ has to be chosen
very carefully, for given values of the other input parameters, in order to
produce ``just a little bit" of $SU(2)$ symmetry breaking. In such a situation
a small decrease of $h_t$ would lead to no symmetry breaking at all ($\mz=0$),
while a slight increase would give \mz\ of order of the SUSY breaking scale.

Barbieri and Giudice \cite{50} required less than 10\% fine--tuning, i.e.
\be \label{e2.29}
\frac {h_t}{m_Z^2} \frac {\partial m_Z^2} {\partial h_t} < 10;
\ee
this leads to bounds on sparticle masses below 1 TeV for squarks and
gluinos, and $\sim 200$ GeV for sleptons and the light chargino. The
quadratic mass dependence means that allowing for fine--tuning at the 1\%
level would ``only" increase these bounds by about a factor of 3.
However, as pointed out in \cite{51}, the bounds become worse once the
1--loop corrections (\ref{e2.15}) to the Higgs potential are included.
As discussed in sec.~2b, the main effect of this correction term can be
absorbed by increasing the scale $Q_0$ where the RG running is terminated,
to a value proportional to the SUSY breaking scale. Since for fixed input
parameters an increase of $Q_0$ decreases the vevs, this reduces the
sensitivity of \mz\ on the SUSY breaking parameters, loosening the upper
bounds on sparticle masses by a about a factor of two.

Unfortunately there is a fair amount of arbitrariness in this procedure.
Obviously the upper bounds will depend on what value one chooses on the r.h.s.
of (\ref{e2.29}). Even worse, one can question whether the constraint
(\ref{e2.29}) corresponds to a true bound on fine--tuning at all. This point
has recently been raised by Anderson and Casta\~no \cite{52}, who pointed out
that (\ref{e2.29}) is a poor measure of our intuitive concept of fine--tuning
if the l.h.s. is large for {\em all} values of the input parameters. In such a
situation no value of \mz\ could be said to be more or less natural than any
other. They therefore suggested to divide the l.h.s. of (\ref{e2.29}) by
something like its average, defined via an integral over parameter space.
Numerically one finds that this modified fine--tuning parameter is about 10
times {\em smaller} than the quantity in (\ref{e2.29}).

While this procedure corresponds more closely to our intuitive concept of
fine--tuning, it also introduces some additional arbitrariness in the result.
The bounds will depend on how exactly one computes the average over the l.h.s.
of (\ref{e2.29}), e.g. what boundaries one uses when integrating over
parameter space. Moreover, in order to derive quantitative results one has to
introduce a probability distribution for the fundamental parameters of the
theory. In their numerical work \cite{52}, Anderson and Casta\~no use two
different distributions, a constant [$f(a)=1$, where $a$ is one of the
fundamental input parameters] and a scale--invariant distribution [$f(a) =
1/a$, which corresponds to a flat distribution on a logarithmic scale]. The
upper bounds on sparticle masses derived using these two choices differ by
typically 10\% or so. Requiring that their fine--tuning parameter be less than
10 then leads to bounds of about 700--800 GeV for squarks and gluinos, 400 GeV
for sleptons, and 250 GeV for the lighter chargino.

There are two fundamental problems besetting all attempts to quantify
fine--tuning: First, one must decide on a probability measure on the parameter
space; second, one must decide how much fine--tuning one can tolerate. Both
these choices are quite subjective. It therefore seems to us that all bounds
on sparticle masses derived from fine--tuning arguments are at best
semi--quantitative. Indeed, it is not clear whether these calculations, which
in case of the method advocated in ref.\cite{52} are quite complicated, are
really an improvement over simply requiring that sparticles should not be
``much heavier than the weak scale", or should be ``at or below the TeV
scale".

Calculations of the LSP relic density seemed to offer a road towards more
precise upper bounds on sparticle masses \cite{40,40a}. We saw in sec.~2c that
requiring $\Omega h^2 \leq 1$ excludes sizable chunks of parameter space even
for relatively modest sparticle masses. Unitarity implies that the LSP
annihilation cross section must decrease like the square of the inverse of the
SUSY breaking scale; the relic density will therefore grow like the square of
that scale, up to logarithmic corrections. It then seems plausible to assume
that this cosmological constraint might provide useful upper bounds on
sparticle masses.

This is indeed true for most values of the dimensionless input parameters,
where we have included the ratios $\mhalf/m_0$ and $A/m_0$ among the
dimensionless parameters of the MSSM. We saw in sec.~2c that the MSSM with
heavy top predicts the LSP to be gaugino--like; if its mass exceeds $m_h/2$ its
annihilation cross section will be dominated by the exchange of sfermions in
the $t-$channel, the exchange of right--handed sleptons being especially
important due to their large hypercharge and relatively small mass. Demanding
that $\Omega h^2 \leq 1$ is then approximately equivalent to the constraint
\cite{40}
\be \label{e2.30}
\frac {\left( m_0^2 + 0.32 m^2_{1/2} \right)^4 }
{ m^2_{1/2} \left( m_0^4 + 0.28 m^2_{1/2} m_0^2 + 0.052 m^4_{1/2} \right) }
\leq 1.8 \cdot 10^5 \ {\rm GeV}^2,
\ee
where the value on the r.h.s. is directly proportional to the bound on
$\Omega h^2$ one assumes. The constraint (\ref{e2.30}) implies \cite{40}:
\ben \label{e2.31} \beq
m_0 & \leq 300 \ {\rm GeV}; \label{e2.31a} \\
\mhalf & \leq 825 \ {\rm GeV}, \label{e2.31b}
\eeq \een
or, in terms of physical sparticle masses:
\ben \label{e2.32} \beq
m_{\tilde{e}_R} &\leq 350 \ {\rm GeV}; \label{e2.32a} \\
m_{\tilde{e}_L} &\leq 630 \ {\rm GeV}; \label{e2.32b} \\
m_{\tilde{W}_1} &\leq 700 \ {\rm GeV}; \label{e2.32c} \\
\mg &\leq 2.0 \ {\rm TeV}; \label{e2.32d} \\
m_{\tilde{q}} &\leq 1.8 \ {\rm TeV}, \label{e2.32e}
\eeq \een
where \msq\ in (\ref{e2.32e}) again refers to first or second generation
squark masses. If these bounds are indeed satisfied, a 1 TeV $e^+e^-$
collider should detect $SU(2)$ singlet sleptons (unless they happen to
lie very close in mass to the LSP), while the LHC should detect squark and
gluino production once ${\cal O}(100)$ fb$^{-1}$ of data have been
accumulated \cite{53}.

Unfortunately these bounds only hold {\em if} the LSP is gaugino--like
{\em and} its annihilation cross section is not ``accidentally" enhanced.
There are several ways of violating at least some of these bounds without
overclosing the universe:
\begin{itemize}
\item If $\mzi \simeq m_h/2$, $s-$channel $h-$exchange can greatly increase
the LSP annihilation cross section, and the bounds (\ref{e2.31a}) and
(\ref{e2.32a}), (\ref{e2.32b}) can be badly violated \cite{42}. In this case
one always has a light chargino, with mass $\mwi \simeq m_h$, and rather
light gluino, with $\mg \simeq 3 m_h$. Some sparticles should therefore be
easily detectable in this scenario.
\item We saw in sec.~2c that the LSP annihilation cross section can be
greatly enhanced if $\mzi > m_t$ and \mst\ is not large. This allows
arbitrarily large values of $m_0$, as long as $\mhalf \leq 750$ GeV so that
\mzi\ is not too much larger than $m_t$. This still leaves the bounds
(\ref{e2.31b}) and (\ref{e2.32c}), (\ref{e2.32d}) more or less unchanged.
However, since (\ref{e2.32e}) can be badly violated, it is not clear whether
gluino production at the LHC would have to be detectable in this scenario;
note that gluinos always decay into $t + \sto$ here. The light stop must not
be too much heavier than the LSP for this scenario to satisfy the relic
density constraint; this potentially very small mass splitting might make
\sto\  difficult to detect at hadron colliders. A decisive test of this
scenario might then have to rely on the bound (\ref{e2.32c}), which
presumably can only be tested at \epem\ colliders with $\rs \geq 1.5$
TeV.
\item As mentioned in secs.~2c and 2d, the LSP could still be higgsino--like
if $\mt(\mt) \leq 155$ GeV and \tanb\ is not small. Such an LSP would have
a very large annihilation cross section into pairs of $W$ and $Z$ bosons
\cite{55,40}. A higgsino--like LSP is therefore cosmologically safe as long
as $\mzi \leq 2$ TeV or so. In this scenario {\em all} sparticles (and all
Higgs bosons except for $h$) could therefore be well beyond the reach of
existing or planned colliders. As long as we strictly stick to our definition
of the MSSM this loophole would be closed if a careful measurement of the top
mass shows that $\mt(\mt) \geq 160$ GeV, i.e. the physical mass exceeds 168
GeV or so. However, it could be re--opened by allowing some very minor
modification of the model, as discussed in sec.~3.
\item We have mentioned several times that the mass of the pseudoscalar Higgs
boson can be greatly reduced if \tanb\ is large; in particular, $m_P \simeq 2
\mzi$ becomes possible, see e.g. table~2. In this case $s-$channel $P-$exchange
contributions will greatly enhance the LSP annihilation cross section, allowing
for gaugino--like LSPs with mass well above 1 TeV.\footnote{Note that
pseudoscalar Higgs exchange proceeds via an $s-$wave; it is therefore more
efficient than scalar Higgs exchange, which contributes only if the two
annihilating LSPs are in a $p-$wave state.}
\end{itemize}

All four scenarios where (some of) the bounds (\ref{e2.31}), (\ref{e2.32}) are
violated involve some amount of fine--tuning, in the sense that some relation
between the input parameters must hold to good precision if the relic density
is to remain acceptably small; this comes on top of the fine--tuning necessary
to ensure that \mz\ is smaller than sparticle masses, as discussed earlier in
this subsection. We nevertheless have to conclude that even within the MSSM we
cannot strictly guarantee the detection of any sparticle at any of the
colliders that are currently being planned. Moreover, it is quite easy to
circumvent the relic density constraint altogether, e.g. by introducing some
amount of $R-$parity breaking which would allow the LSP to decay into SM
particles; this possibility is discussed in sec.~4e.

Of course, all this does not mean that we actually expect the LSP mass to
exceed 1 TeV; rather, we feel that SUSY would lose much of its appeal if
the LHC or a TeV \epem\ collider failed to detect any sparticles. However,
one should keep in mind that even within the MSSM this sentiment cannot be
backed up by mathematically strict upper bounds on sparticle masses.

\setcounter{footnote}{0}
\section*{3) Minor Modifications of the MSSM}
In this section we discuss extensions or modifications of the MSSM that do
not involve new light (s)particles or new interactions between MSSM
particles. In sec.~3a we discuss sum rules that allow one to test various
aspects of SUSY breaking quantitatively, and in sec.~3b we describe how the
existence of a GUT sector might modify the sparticle spectrum

\subsection*{3a) Testing SUSY Breaking via Sum Rules}
The boundary conditions (\ref{e2.3}) describing SUSY breaking at the GUT scale
are important ingredients of our definition of the MSSM. They led to
eqs.(\ref{e2.16}) for first and second generation sfermion masses, and
eq.(\ref{e2.18}) for gaugino masses. These latter equations obviously
allow to test the boundary conditions once all sparticle masses are
known. However, it is often more convenient to test specific relations
(``sum rules") between sparticle masses which are independent of the
input parameters at the GUT scale, and which in principle allow to test
different aspects of SUSY breaking separately.

An obvious, yet very important example is the near--degeneracy of sfermions
of the first two generations with identical gauge quantum numbers,
$m_{\tilde{e}_R} = m_{\tilde{\mu}_R}, \ m_{\tilde{u}_L} = m_{\tilde{c}_L}$,
and so on. This remains true as long as there are no new large Yukawa
interactions involving the first two generations. In particular, one need
not assume eqs.(\ref{e2.3a}) or (\ref{e2.3c}). Extensions of the gauge group
will also leave this result unchanged as long as the first two generations
couple with equal strength to the new gauge boson(s); this is true, e.g., for
popular extensions involving new $U(1)$ factors and/or an $SU(2)_R$
symmetry.

The following sum rules are of even more general validity:
\be \label{e3.1}
m^2_{\tilde{d}_L} - m^2_{\tilde{u}_L} = m^2_{\tilde{e}_L} - m^2_{\tilde{\nu}}
= - M^2_W \cos(2 \beta).
\ee
They follow directly from $SU(2)_L$ invariance as long as the current
eigenstates are also mass eigenstates. Given that $SU(2)_L$ invariance in
the matter sector has been tested quite thoroughly, a violation of one of
these relations would most likely indicate mixing between different current
states. However, we know that mixing between $SU(2)$ singlet and doublet
sfermions of the same electric charge has to be small for the first two
generations, due to the smallness of the corresponding fermion masses, see
eq.(\ref{e2.12}). Mixing with exotic sfermion states would presumably also be
tightly constrained by the strong bounds \cite{56} on mixing of SM fermions
with
exotic states. We therefore consider eq.(\ref{e3.1}) to be the by far most
reliable theoretical prediction involving sparticle masses.

The following sum rule is slightly more model dependent \cite{57}:
\be \label{e3.2}
2 \left( m^2_{\tilde{u}_R} - m^2_{\tilde{d}_R} \right)
+ \left( m^2_{\tilde{d}_R} - m^2_{\tilde{d}_L} \right)
+ \left( m^2_{\tilde{e}_L} - m^2_{\tilde{e}_R} \right)
= \frac{10}{3} \stw M^2_Z \cos(2\beta).
\ee
It relies on the assumption that sfermions within one generation are degenerate
at some energy scale.\footnote{More exactly, it only assumes that the l.h.s.
vanishes at some scale (up to $D-$terms), since the combination of explicit
SUSY breaking mass terms that appears here is an RG invariant in the absence of
new interactions.} It does not assume degeneracy of gaugino masses at any
scale.
However, unlike the degeneracy of sfermions with identical gauge quantum
numbers it will in general not be valid in the presence of new gauge
interactions.

It is difficult to derive useful sum rules for third generation sfermion masses
if \tanb\ is very large, in which case all third generation Yukawa couplings
are sizable. However, in the important special case where only $h_t$ affects
the RG running of sfermion masses significantly additional sum rules can be
given. Within the MSSM they should hold to good precision as long as
$\tanb \leq 15$ or so. In particular \cite{57},
\be \label{e3.3}
m^2_{\tilde{t}_1} + m^2_{\tilde{t}_2} - 3 m^2_{\tilde{b}_L} - 2 m^2_t =
m^2_{\tilde{u}_L} + m^2_{\tilde{u}_R} - 3 m^2_{\tilde{d}_L};
\ee
recall that $\tilde{b}_L$ is to good approximation a mass eigenstate if
\tanb\ is not large. This sum rule tests degeneracy of sfermions both
within one generation and between the first two and the third generation; it
is again independent of gaugino masses. The same assumptions also imply
\be \label{e3.4}
\left( m^2_{\tilde{t}_2} - m^2_{\tilde{t}_1} \right)^2 = 4 m^2_t
\left( A_t + \mu \cot \! \beta \right)^2 +
\left( m^2_{\tilde{b}_L} - m^2_{\tilde{d}_L} - m^2_{\tilde{u}_L}
+ m^2_{\tilde{u}_R} \right)^2,
\ee
which allows to determine $|A_t + \mu \cot \! \beta|$ from the
knowledge of sparticle masses alone, {\em if} sfermion masses are
unified at some scale.

None of the sum rules we have discussed so far tests the unification of
gaugino masses. It is in principle possible to test eq.(\ref{e2.3a})
using sfermion masses \cite{57}, at least if eq.(\ref{e2.3b}) holds; in
practice this is hampered by our at present poor knowledge of the
coefficient \cg\ appearing in eqs.(\ref{e2.16}). It is therefore probably
preferable to test relations between gaugino masses more directly. In
particular, it has been shown \cite{58} that the relation
\be \label{e3.5}
M_1 = \frac{5}{3} \tan^2 \theta_W M_2 \simeq 0.5 M_2
\ee
can be tested to good precision using measurements of chargino production
cross sections at an \epem\ collider with polarized beams. An even more
direct test of the unification of gaugino masses can be made using the
following sum rule \cite{57}
\be \label{e3.6}
\sum_{i=1}^4 \epsilon_i |m_{\tilde{Z}_i}| = \frac {\alpha_1 + \alpha_2}
{\alpha_3} M_3.
\ee
Here $\epsilon_i$ is the sign of the $i.-$th eigenvalue of the neutralino
mass matrix. In most of the allowed parameter space only one $\epsilon_i$ is
negative; in the MSSM this will be one of the higgsino--like states. The
$\epsilon_i$ can e.g. be determined from an analysis of off--diagonal
$\wt{Z}_i \wt{Z}_j$ production at \epem\ colliders. The $\alpha_i$ are the
fine structure constants of the three factor groups of the SM, with $\alpha_1$
in a GUT normalization. Finally, $M_3$ is the running gluino mass; it is
related to the on--shell mass by eq.(\ref{e2.19}). Unfortunately this means
that for a precision test of eq.(\ref{e3.6}) we also have to know the squark
masses.

Finally, under the same assumptions that led to eqs.(\ref{e3.4}) and
(\ref{e3.5}) one has:
\be \label{e3.7}
m_P^2 = \frac { m^2_{\tilde{\nu}} + \mu^2 } { \sin^2 \beta };
\ee
this is simply eq.(\ref{e2.28}) for small or moderate \tanb. This sum
rule tests whether at scale \mx\ SUSY breaking scalar masses are the same
in the Higgs and matter ($\tilde{\nu}$) sectors; it is again independent of
the unification of gaugino masses. Unfortunately the r.h.s. of eq.(\ref{e3.7})
cannot easily be written in terms of physical sparticle masses alone. If
$|\mu|$ is large, it is approximately equal to the masses of the heavy
neutralino and chargino states \cite{40,57}. More constraints on $\mu$ and
$\beta$ can be derived from the following relations \cite{57}:
\ben \label{e3.8} \beq
m^2_{\wt{W}_1} + m^2_{\wt{W}_2} &= M_2^2 + \mu^2 + 2 M_W^2 ;
\label{e3.8a} \\
\mwi m_{\wt{W}_2} &= \mu M_2 - M_W^2 \sin \ 2 \beta,
\label{e3.8b} \eeq \een
which follow directly from the structure of the chargino mass matrix and thus
only rely on the assumption of minimal particle content. Notice that
eq.(\ref{e3.8b}) is more sensitive to $\beta$ than the sfermion masses if
$\tanb \geq 3$ or so.

The sum rules presented in this subsection hold at the RG--improved tree
level. With the exception of eq.(\ref{e3.6}) we expect radiative corrections
to these relations to be quite small, although they obviously will have to be
included \cite{59} if these sum rules are to be tested at the percent level.
Many of these relations involve differences of large numbers, and necessitate
light flavor tagging; it therefore seems unlikely that good tests can be
performed using hadron collider data only, although knowledge of the gluino
mass might well have to come from hadron collider experiments.

\subsection*{3b) GUT Effects}
Broadly speaking the presence of a GUT sector can change predictions for
the sparticle spectrum in two ways: It can lead to relations between
parameters that would otherwise be independent of each other; and it
can change the sparticle spectrum at scale \mx, i.e. the boundary
conditions (\ref{e2.3}). The first effect usually increases the predictive
power of the model, while the second tends to decrease it since it will
depend on the parameters of the GUT model. We will discuss both kinds of
effects in turn.

\subsubsection*{3b1) GUT--Imposed Constraints}
The most studied \cite{60,8} additional prediction due to GUTs is based on the
relation $h_b(\mx) = h_{\tau}(\mx)$, which holds in $SU(5)-$like GUTs (up to
small threshold corrections). As shown in fig.~6, taken from Barger et al.
\cite{8}, this bottom--tau Yukawa unification only leads to the correct value
of $m_b$ (assuming $m_{\tau}=1.78$ GeV) if either $h_t$ is very large
(fixed--point solution), or $\tanb \simeq \mt(\mt)/m_b(\mt)$, in which case
$h_b$ is quite large. If all Yukawa couplings were much smaller than the
strong coupling, QCD effects would increase $h_b$ too much, giving $m_b(m_b)
\sim 5-5.5$ GeV; a large $h_t$ or large $h_b$ can slow down the running of
$h_b$ as required by the experimental result $m_b(m_b) = 4.25 \pm 0.1$ GeV.
The parameter space of the MSSM with large $h_t$ was explored in sec.~2c, and
in sec.~2e we discussed the case of large \tanb.\footnote{It has recently been
pointed out \cite{61} that for $\tanb \gg 1$, weak--scale threshold
corrections to $m_b$ can be large; this changes the shape of the allowed
region with large \tanb, depending on the values of the soft--breaking
parameters. However, it still remains true that bottom--tau Yukawa unification
requires either a very large $h_t$ or very large \tanb.}

In $SO(10)$ models where the weak--scale Higgs doublets reside entirely
within a {\bf 10}--dimensional representation, one expects all three Yukawa
couplings of the third generation to be unified; this uniquely singles out
solutions with rather heavy top quark and very large \tanb\ \cite{62,63}.

\makebox[15.5cm]{\epsfxsize=0.67\hsize\epsffile{sumo6.ps}}

\noindent {\bf Fig.6:} {\small The region of the $(\mt, \tanb)$ plane that is
compatible with the unification of the bottom and tau Yukawa couplings at the
GUT scale, for various combinations of $\alpha_S(\mz)$ and the effective SUSY
threshold. The recently discovered \cite{61} weak--scale threshold corrections
to $m_b$ have not been included here; they might be significant for very large
\tanb. Adapted from Barger et al. \cite{8}.}\\
\vspace*{4mm}

It should be mentioned that not even $b-\tau$ unification can be extended
successfully to the first and second generation. Instead one often assumes
that the masses of these light fermions are created by non--renormalizable
(effective) operators, and are thus suppressed by powers of $M/\mpl$, where
$M$ is some mass of order \mx. One can construct realistic fermion mass
matrices in this fashion, often with fewer parameters than a completely
general ansatz would have \cite{63,64}. Unfortunately it is not easy to test
such schemes, since they usually allow some variation in the predicted
values of masses and CKM angles. One possibility is to extend the ansatz
to the neutrino sector where little solid experimental information exists as
yet, so that genuine predictions (as opposed to fits of data) are still
possible. None of the schemes suggested to reproduce first and second
generation fermion masses seems to lead to new predictions for the sparticle
or Higgs spectrum.

A more direct test of the existence of a GUT would be the detection of a
process which is strictly forbidden within the MSSM. The most widely studied
example \cite{65,9} is nucleon decay. Unfortunately predictions for the
lifetime of the proton depend quite strongly on details of the GUT model. In
minimal SUSY $SU(5)$ and similar models the largest contributions to nucleon
decay amplitudes involve the exchange of the colored $SU(5)$ partner of the
weak--scale $SU(2)$ doublet higgsinos. Within minimal $SU(5)$ one can
determine the masses of these $SU(3)$ triplet higgsinos, with some
uncertainty, from their threshold effects on the running of the gauge
couplings \cite{66,15b}. The result tends to come out rather low; the existing
experimental bounds on proton decay then lead to rather strong constraints on
the parameter space of the MSSM \cite{9,42}. In particular, solutions with
large \tanb\ are disfavoured since some of the Yukawa couplings appearing in
nucleon decay amplitudes grow $\propto \tanb$. On the other hand, a large
value of $m_0/\mhalf$ tends to suppress nucleon decay, since higgsino exchange
leads to a virtual two--sfermion intermediate state which has to be
transformed into a two--fermion state by the exchange of a weak--scale
chargino or neutralino; the resulting ``dressing loop function" is $\propto
\mhalf/m^2_0$. Including bottom--tau unification, minimal SUSY $SU(5)$ thus
predicts small \tanb, large $h_t$, and $m^2_0 \gg m^2_{1/2}$; the resulting
phenomenology has been studied in ref.\cite{47}.

However, minimal SUSY $SU(5)$ is not really a satisfactory model, since the
required huge mass splitting between the weak--scale Higgs doublets and their
$SU(5)$ partners requires extreme fine--tuning of parameters in the GUT
superpotential. This is still some improvement over the non--supersymmetric
version since this hierarchy, once created, will be stable against radiative
corrections \cite{3}; nevertheless it is certainly not a very appealing
solution of the hierarchy problem. More elegant solutions \cite{67} require
a more complicated GUT Higgs sector. In such models the higgsino exchange
contributions to nucleon decay are often severely suppressed or altogether
absent; these models do therefore not impose additional restrictions on the
values of the soft breaking parameters of the MSSM.

It has recently been argued by Barbieri and Hall \cite{68} that predictions for
lepton--number violating processes \cite{69} can be made much more reliably in
SUSY GUTs than predictions for nucleon decay. The basic observation is that any
GUT model worth its name will unify (s)quarks and (s)leptons within a single
representation of the GUT gauge group. We know experimentally that there is
mixing between quarks of different generations; this means that the
corresponding Yukawa couplings must be nontrivial matrices in generation space.
The non--diagonal Yukawa couplings will then generate non--diagonal entries in
the slepton mass matrices at the one loop level. These non--diagonal entries
will be left essentially unchanged by the RG running from \mx\ down to the
weak scale. The net effect is that in SUSY GUTs, lepton flavour violating
processes like $\mu \rightarrow e \gamma$ are only suppressed by powers of
slepton masses; this is in sharp contrast to non--SUSY GUTs where, in the
absence of neutrino masses, these processes are suppressed by powers of \mx\
and are thus completely unobservable. Barbieri and Hall estimate \cite{68}
that for minimal SUSY $SU(5)$ with large $h_t$, $\mu \rightarrow e \gamma$
decays and $\mu \rightarrow e$ conversion in matter should occur at rates
about ten times below present bounds for slepton masses of 100 GeV; the rates
scale like $m^{-4}_{\tilde l}$. More importantly, they argue that this
estimate remains roughly valid in a wide class of models, including those
giving rise to realistic quark and fermion mass matrices. At present this
does not lead to new bounds on sparticle masses beyond those derived from
LEP searches, but it certainly gives renewed impetus to efforts to improve
the sensitivity of searches for lepton--flavour violating processes.

As a final example for additional constraints on parameter space due to
the presence of a GUT we mention the result of ref.\cite{70}, where the
bound $|B(\mx)| \geq 2 m_0$ was derived under the assumption that the
$\mu$ parameter is created entirely by integrating out superheavy fields of
the GUT sector; recall that $B$ appears in the low energy Higgs potential
(\ref{e2.4}). This mechanism offers an appealing solution of the $\mu$ problem
(see Sec.~2.1), since the value of $\mu$ created in this way is automatically
of order of the SUSY breaking scale. However, other solutions to the $\mu$
problem have been suggested \cite{71,71a}.

\subsubsection*{3b2) Changes of the Boundary Conditions}
In a supergravity context one would expect the boundary conditions
(\ref{e2.3}) to hold at the Planck scale rather than the GUT scale. RG
running between \mpl\ and \mx\ can then change some sparticle and Higgs
masses significantly. Additional contributions can come from GUT scale
threshold effects (at one loop level), or due to $D-$terms if the rank of
the GUT group is bigger than four.

The possible importance of the RG scaling of soft breaking parameters
between \mpl\ and \mx\ was recognized about ten years ago \cite{72}.
More recently this has been studied in some detail in ref.\cite{73}, within
the framework of minimal $SU(5)$. The relevant terms in the superpotential
are:
\be \label{e3.9}
W = \lambda' {\rm tr} \Sigma^3 + M_{\Sigma} {\rm tr} \Sigma^2 + M_H H_1 H_2
+ \lambda H_1 \Sigma H_2 + h_U \Psi \Psi H_2 + h_D \Psi \Phi H_1.
\ee
Here $\Sigma$ denotes a {\bf 24} and breaks $SU(5)$, $H_1$ and $H_2$ are ${\bf
\bar{5}}$ and {\bf 5} representations containing the light Higgs fields, and
$\Psi(10)$ and $\Phi(\bar{5})$ contain the matter superfields. The RG running
between \mpl\ and \mx\ then has essentially three effects:
\begin{itemize}
\item We know that at least one Yukawa coupling in eq.(\ref{e3.9}), that
of the top quark, is quite big. The $h_U$ term will therefore tend to
reduce $m^2_{H_2}$ at \mx, and also the masses of the third generation
sfermions residing in the {\bf 10}, i.e. $\tilde{t}_L, \ \tilde{t}_R, \
\tilde{b}_L$ and $\wt{\tau}_R$. Apart from the reduction of
$m^2_{\tilde{\tau}_R}$ this merely enhances trends that already exist in
the MSSM. In particular, the reduction of $m^2_{H_2}$ increases the value
of $|\mu|$ determined from weak symmetry breaking even further.
\item If the coupling $\lambda$ in eq.(\ref{e3.9}) is large, it will
reduce the masses of both $H_1$ and $H_2$ at scale \mx, compared to
squark and slepton masses. This increases the value of $|\mu|$ yet
again, but leaves the masses of the physical Higgs fields at the weak
scale more or less unchanged.
\item Gaugino loops will increase the masses of members of {\bf 10}--plets
over those of {\bf 5}--plets. In particular, this increases the masses
of stop squarks compared to those of Higgs bosons, and therefore once
again requires an increase of $|\mu|$ to get correct symmetry breaking at
the weak scale. Notice that the right--handed (s)leptons also reside in the
{\bf 10}. As a result, in SUSY $SU(5)$ one has \cite{74}
\be \label{e3.10}
m^2_{\tilde{e}_R} \geq 3.1 M_1^2,
\ee
where $M_1$ is the bino mass; this is to be compared with $m^2_{\tilde{e}_R}
\geq 0.87 M_1^2$ in the MSSM. This new constraint reduces the annihilation
cross section of bino--like LSPs for given values of $m_0$ and \mhalf, and
therefore {\em lowers} the bounds on sparticle masses that follow from
imposing an upper limit on the LSP relic density in the ``generic" case of a
bino--like LSP whose annihilation cross section is not enhanced by $s-$channel
contributions. In particular, the coefficients of $m^2_{1/2}$ and $m^4_{1/2}$
in the inequality (\ref{e2.30}) change, which implies:
\ben \label{e3.11} \beq
m_0 &\leq 225 \ {\rm GeV}; \label{e3.11a} \\
\mhalf &\leq 400 \ {\rm GeV}. \label{e3.11b}
\eeq \een
In particular (\ref{e3.11b}) is a much tighter constraint than (\ref{e2.31b});
it implies $\mg \leq 1$ TeV, so that gluino production should be quite easily
observable at the LHC. Recall, however, that these bounds have several
loopholes, as discussed in sec.~2f.
\end{itemize}

Altogether the authors of ref.\cite{73} found that, at least within minimal
$SU(5)$, RG scaling between \mpl\ and \mx\ increases $|\mu|$, often quite
substantially; the masses of the heavy Higgs bosons are usually also increased,
although by a smaller amount. For the same set of input parameters the mass
of the lightest stop is also increased; however, very small values of \mst\
are not excluded, since combinations of parameters that previously gave
$m^2_{\tilde{t}_1} < 0$ might now be allowed.

An additional complication arises when the rank of the GUT group is bigger
than four. In this case at least one diagonal group generator needs to be
broken to reach the SM gauge group; in general this will lead to the
occurrence of nonzero $D-$term contributions to scalar masses \cite{75,74}.
In an $SO(10)$ model they might allow small values of $|\mu|$ by giving
positive contributions to $m^2_{H_2}$ \cite{74}. On the other hand, if the
$D-$term contribution is negative for $m^2_{H_2}$ and positive for
$m_{H_1}^2$, it becomes easier \cite{76} to find solutions with $\tanb
\simeq m_t/m_b$, since even if $h_t(\mx) = h_b(\mx)$ one has $m^2_{H_1}
> m^2_{H_2}$ at the weak scale, as required for $\tanb > 1$. The sign of
the $D-$terms depends on the details of the GUT model. The breaking of
additional $U(1)$ factors at an intermediate scale will in general also
produce $D-$term contributions to scalar masses \cite{75}; these contributions
are therefore expected to exist in all models where the rank of the gauge
group exceeds four. Such terms could, e.g., destroy the near--degeneracy
of squarks within the first or second generation. Since they do not lead to
mass splitting between squarks with equal gauge quantum numbers these
$D-$terms do not create problems with flavor--changing neutral currents
(FCNC); nevertheless it is clear that sizable mass splittings between
different squarks can have substantial impact on collider phenomenology.

One of the most vexing problems in SUSY model building is the failure to
construct a realistic GUT model based on the superstring \cite{77}. Part of the
problem is that most GUT models need rather large Higgs representations; these
can only be produced in string models if one goes to higher levels of the
underlying Kac--Moody algebra, which poses technical difficulties. The
semi--realistic models found in ref.\cite{77} usually predict the existence of
several new light fields, which destroyed the successful prediction for the
gauge couplings at low energies. The ``flipped SU(5)" model of ref.\cite{78}
manages to break the underlying $SU(5) \times U(1)$ group down to the SM gauge
group using only {\bf 10}--dimensional Higgs representations. In addition it
leads to automatic mass splitting between the light doublet Higgs fields and
their GUT partners, and uniquely singles out $SU(3) \times SU(2) \times U(1)$
as low energy gauge group. Finally, higgsino exchange does not contribute to
nucleon decay in this model. Unfortunately it is not fully unified, since part
of the $U(1)_Y$ factor of the SM gauge group resides in the additional $U(1)$
factor of the GUT group. Superstring theory predicts the two in principle
independent couplings of the model to unify at the string scale $M_C \simeq 5
\cdot 10^{17}$ GeV; however, this gives the correct values for the gauge
couplings at low energy only if one introduces additional fields below the
unification scale. In this model the apparent unification of gauge couplings
with MSSM field content is thus a mere accident. On the positive side, it does
not appear to be difficult to derive flipped $SU(5)$ models from superstring
theory. See ref.\cite{78} for further discussions of the phenomenology of this
model.

We already mentioned that string theory predicts unification of all gauge
interactions even if there is no real GUT sector. Unfortunately the predicted
unification scale is too high ($5 \cdot 10^{17}$ GeV), unless there are large
threshold corrections. While the existence of such large corrections cannot be
excluded, from the string point of view they appear to be somewhat unexpected
\cite{79}. The problem can be solved by the introduction of additional states
below \mx, as in the flipped $SU(5)$ model, but then one merely fits, rather
than predicts, the low energy gauge couplings.

\setcounter{footnote}{0}
\section*{4) Major Modifications of the MSSM}
In this section we discuss what we consider to be major modifications
of the MSSM. We will be quite brief here, partly for reasons of space,
and partly because not too many significant predictions for or
constraints on sparticle or Higgs masses are known within these models,
which almost always contain (far) more parameters than the MSSM does.
However, we will try to provide the interested reader with references where
more details can be found. Specifically, in sec.~4a we discuss what happens
if the boundary conditions (\ref{e2.3}) are given up, still keeping a
``Grand Desert" scenario. We then discuss models with extended Higgs (4b),
gauge (4c) and matter (4d) sectors. In sec.~4e we briefly describe models
with broken $R-$parity, and sec.~4f is devoted to a discussion of recent
attempts to break SUSY dynamically at a rather low scale (compared to \mpl).

\subsection*{4a) Arbitrary SUSY Breaking at the Planck Scale}
We have already seen in sec.~3b that the MSSM predictions of sec.~2 can be
altered quite significantly if we modify the boundary conditions (\ref{e2.3}),
e.g. due to RG scaling between \mpl\ and \mx. In this subsection we address the
question whether one can derive significant constraints on the sparticle or
Higgs spectrum simply from the assumption of a Grand Desert (no intermediate
scale between the SUSY breaking and GUT or Planck scales), without having to
specify the boundary conditions at very high energies.\footnote{Of course,
constraints from direct experimental searches are more or less independent
{}from assumptions on the sparticle spectrum, at least as far as searches at
\epem\ colliders are concerned. Here we are mostly interested in constraints
imposed by the structure of the model itself, similar to our discussion in
sec.~2.}

Not surprisingly, shifts of the allowed region of parameter space are usually
not very large if deviations from universality at the GUT scale are modest
\cite{81}. The region of very large \tanb\ is an exception to this rule, since
it is quite sensitive to the ratio of Higgs masses at the GUT scale, as already
discussed in sec.~3b. Such minor deviations might e.g. be expected in
supergravity models with almost (but not entirely) flat K\"ahler metric.
However, if we do not impose any constraints on this metric (or, equivalently,
the form of the kinetic energy terms of chiral superfields near \mpl),
arbitrary scalar masses can be generated already at the Planck scale
\cite{82,74}; not even the presence of a GUT sector necessarily implies any
relations between scalar masses (between members of the same GUT group
multiplet, for example), if the kinetic terms for the light chiral
superfields depend on the fields that break the GUT group \cite{83}. Large
differences between scalar masses can also appear easily in certain string
models \cite{80}; this has spawned renewed interest in this kind of generalized
models.

Obviously the number of free parameters is increased greatly if we do not
impose constraints on the scalar spectrum at some very high energy; an
exhaustive scan of parameter space therefore becomes all but impossible. A
perhaps more serious criticism is that models where the two $SU(2)$ doublet
Higgs bosons have different masses already at the Planck scale do in general
not need any radiative corrections in order to achieve weak gauge symmetry
breaking. Note that attempts \cite{84} to understand the size of soft breaking
masses dynamically within supergravity or string models all rely on radiative
gauge symmetry breaking. The basic observation is that radiative symmetry
breaking requires $\log(\mpl/M_{\rm SUSY}) \gg 1$. If the SUSY breaking scale
is somehow related dynamically to the vevs of the weak Higgs doublets it is
therefore naturally expected to be exponentially smaller than the Planck
scale, as required if SUSY is to solve the hierarchy problem. In contrast, the
application of this idea to models where Higgs fields acquire a vev already at
tree level would give an ``infinite" SUSY breaking scale. However, this can be
used only as an argument in favor of (nearly) equal Higgs masses at the GUT
scale; it is independent of the form of the squark and slepton spectrum.

Certain inequalities between sparticle masses will still hold if we assume
that all squared squark and slepton masses are positive (or zero) at the
GUT scale. In particular, this implies \cite{30} that squarks of the first
and second generation cannot be much lighter than gluinos, see
eq.(\ref{e2.21}).
It should be noted, however, that there is nothing intrinsically wrong with
having $m^2_{\tilde{q}} < 0$ at scale \mx\ as long as $|m_{\tilde{q}}|$ is
of order of the weak scale. In this case the tree--level potential renormalized
at scale \mx\ will have a minimum with $\langle \tilde{q} \rangle$ of order
of the weak scale, or possibly of order $|m_{\tilde{q}}|/h_q$ where $h_q$ is a
Yukawa coupling. However, in order to determine whether this minimum really
exists one should minimize the potential renormalized at the scale of this vev
\cite{22}. If the running of $m^2_{\tilde{q}}$ between \mx\ and this lower
scale increases it to a positive value, this minimum is in fact spurious. In
this case the inequality (\ref{e2.21}) might be badly violated.\footnote{This
mechanism will not work in the MSSM since there the masses of $SU(2)$ singlet
sleptons run very little between \mx\ and \mz. The requirement $m_{\tilde{e}_R}
> \mzi$ then excludes the possibility that $m_0^2 \ (=m^2_{\tilde{q}}$ at scale
$\mx)$ is significantly less than zero. Of course, this argument breaks down if
we allow squark and slepton masses to differ already at the GUT scale.}

It is often claimed that a high degree of degeneracy at least between certain
squark masses is needed in order to avoid problems with flavor changing neutral
current (FCNC) processes, e.g. in \kkbar\ mixing. The reason is that the
introduction of completely arbitrary squark mass matrices (in generation space)
in general gives rise to large flavor off--diagonal $\gl \tilde{q}_i q_j$
couplings, where $\tilde{q}_i$ and $q_j$ are mass eigenstates. Since this
involves strong interactions, the size of such couplings is indeed constrained
quite severely \cite{85}. These constraints are most easily derived in the
basis where the $\gl \tilde{q}_i q_j$ couplings are diagonal but the squark
mass matrices are not; one can then give rather strong bounds on the
off--diagonal entries of these matrices.

For example, from expressions given by Gabbiani and Masiero \cite{85} one can
derive the bound
\be \label{e4.1}
\Delta_{\tilde{d} \tilde{s}} \leq 1.3 \cdot 10^{-5} \ {\rm GeV}^{-1} \cdot
\overline{m}^3_{\tilde{d} \tilde{s}}
\ee
{}from the requirement that $\gl-\tilde{q}$ box diagrams do not contribute more
than the experimental value of $3.5 \cdot 10^{-15}$ GeV to the $K_L-K_S$ mass
difference. Here $ \Delta_{\tilde{d} \tilde{s}}$ is the off--diagonal (1,2)
entry in the mass matrix for charge $=1/3$ squarks and
$\overline{m}^2_{\tilde{d} \tilde{s}}$ is the geometric mean of the diagonal
(1,1) and (2,2) entries. In deriving the bound (\ref{e4.1}) we have assumed
identical mass matrices for $SU(2)$ doublet and singlet squarks and have
ignored mixing between these two sectors. If only one of these two mass
matrices has an off--diagonal (1,2) entry the coefficient in (\ref{e4.1})
has to be increased to $4.4 \cdot 10^{-5}$ GeV$^{-1}$. (These coefficients
depend weakly on the ratio $\mg/\msq$; the given numbers are for $\mg=\msq$.)

It should be noted, however, that (\ref{e4.1}) is not really a bound on squark
mass splitting, i.e. on the difference of the eigenvalues of the squark mass
matrices. For example, if these matrices are diagonal in the current basis, one
has $\Delta_{\tilde{d}\tilde{s}} = \sin \! \theta_C \cos \! \theta_C \left(
m^2_{\tilde{d}} - m^2_{\tilde{s}} \right) \simeq \left( m^2_{\tilde{d}} -
m^2_{\tilde{s}} \right)/5$. This still gives quite a strong bound on
$\tilde{d}- \tilde{s}$ mass splitting, requiring $|m_{\tilde{d}} -
m_{\tilde{s}}| \leq 1.5 \ (35)$ GeV for  $\overline{m}_{\tilde{d} \tilde{s}} =
0.2 \ (1.0)$ TeV. However, due to the smallness of the mixing angles between
the third and the first two generations of quarks, quite sizable mass
splittings are allowed between squarks of the first two and the third
generation {\em if} the squark mass matrix is flavor--diagonal in the current
basis; this is why the MSSM is not much constrained by FCNC processes even if
the top quark is heavy, i.e. third generation squark masses are reduced
substantially.

Finally, it is possible that the squark mass matrix is (almost) perfectly
aligned with the quark mass matrix, e.g. due to a horizontal symmetry
\cite{86}. In this case gluino and neutralino exchange do not contribute at
all to FCNC processes. Squark mass splitting is then constrained only by
bounds on chargino exchange contributions, which vanish identically for
degenerate squarks due to the super--GIM mechanism \cite{85a}. Since we are
now dealing with a purely weak process which in addition is suppressed by
CKM mixing elements the resulting bounds are much weaker than in the previous
case. For example, the bound from \kkbar\ mixing gives:
\be \label{e4.2}
\left| \frac {2 \left( m^2_{\tilde{u}_L} - m^2_{\tilde{c}_L} \right) }
{m^2_{\tilde{u}_L} + m^2_{\tilde{c}_L} } \right| \leq \frac {\mwi}
{1 \ {\rm TeV}} \sqrt{ 2 + \frac { m^2_{\tilde{u}_L} + m^2_{\tilde{c}_L} }
{ 8 m^2_{\tilde{W}} } }.
\ee
A slightly weaker bound on the $\tilde{d}_L-\tilde{s}_L$ mass difference
follows from the experimental bound on $D^0 - \overline{D^0}$ mixing.
Notice that squark mixing in the $SU(2)$ singlet squark sector is now
completely unconstrained. Moreover, the bound (\ref{e4.2}) is for on--shell
squark masses. The bounds on the running squark masses at the GUT or Planck
scale are usually much weaker, due to the large contribution from gluino loops
which is, of course, the same for all flavors.\footnote{The running of flavor
off--diagonal entries of squark mass matrices and soft breaking $A$ parameters
has recently been studied in ref.\cite{85b}.} For example, (\ref{e4.2})
translates into the following constraint on GUT--scale quantities:
\be \label{e4.3}
\left| \delta_{\tilde{u}\tilde{c}} \right| \leq \frac {0.8 \mhalf}
{1 \ {\rm TeV}} \left( 1 + 6 \frac {m^2_{1/2}} {\overline{m}^2_0} \right)
\sqrt { 4.3 + 0.4 \frac {\overline{m}^2_0} {m^2_{1/2}} },
\ee
where $\overline{m}^2_0$ is the average and $\overline{m}^2_0
\delta_{\tilde{u}\tilde{c}} $ the difference between squared $\tilde{u}_L$ and
$\tilde{c}_L$ squark masses at scale \mx. Notice that
$\delta_{\tilde{u}\tilde{c}} \simeq 1$ is allowed already for $\overline{m}_0
= \mhalf = 100 $ GeV.

We therefore conclude that experimental constraints on FCNC processes do lead
to significant bounds on {\em mixing} between squarks of the first two
generations; in ``generic" models, e.g. if the squark mass matrices are
diagonal in the current basis, the mass splitting between first and second
generation squarks of the same charge is then also tightly constrained.
However, much larger mass splittings are allowed for the third generation;
{\em no} bound on mass splitting between squarks of different charge exists;
and even for squarks with the same charge the bounds become quite weak if
squark and quark mass matrices are aligned, e.g. due to a horizontal symmetry.
Moreover, there could be ``accidental" cancellations between different SUSY
contributions; this possibility has not been included when deriving the
constraints (\ref{e4.1}), (\ref{e4.2}). Hence the assumption of a Grand Desert
scenario, even when combined with experimental bounds on FCNC processes, does
not lead to constraints on sparticle masses that are of general validity,
apart from the sum rule (\ref{e3.1}) which follows directly from $SU(2)$
invariance.

\subsection*{4b) Models with Extended Higgs Sector}
One of the most frequently studied extensions of the MSSM is the model with
an additional Higgs singlet superfield $N$. This model has the virtue of
allowing a purely cubic superpotential; there is no need to introduce a
supersymmetric mass term ($\mu-$term). Instead, one introduces Higgs
self--couplings in the superpotential:
\be \label{e4.4}
W_N = \lambda N H_1 H_2 + \lambda' N^3,
\ee
so that $\lambda \langle N \rangle$ serves as an effective $\mu-$term. Since
$N$ is a gauge singlet, the running of the gauge couplings is not affected
significantly; the unification of the gauge couplings observed in the MSSM
therefore carries over to this model. However, the introduction of a gauge
singlet might be problematic. If it couples to superheavy GUT sector fields,
the stability of the gauge hierarchy cannot be guaranteed at the quantum level
\cite{87}.

The most important phenomenological difference to the MSSM results from the
fact that we now have introduced Higgs self--interactions of unknown strength,
in addition to the gauge interactions given by the $D-$term. As a result, we
can give an upper bound on the mass of the lightest neutral Higgs scalar only
if we can bound the coupling $\lambda$ in eq.(\ref{e4.4}) from above; in this
respect the model resembles the nonsupersymmetric SM. Such a bound can be
derived \cite{89} from the requirement that $\lambda$ remains perturbative up
to scales of order \mx. More recently it has been shown \cite{90} that the
bound on $\lambda$ becomes stronger as the Yukawa coupling of the top quark
increases. As a result, the upper bound on the mass of the lightest Higgs
boson depends only rather weakly on \mt\ once Coleman--Weinberg corrections to
the scalar potential are included, which grow $\propto m_t^4$ as in the MSSM.
Numerically the bound amounts to about 150 GeV.

Note that the physical Higgs spectrum of this model contains three neutral
scalars and two pseudoscalars. The lightest Higgs eigenstate could now be
dominantly a gauge singlet, in which case it couples very weakly to gauge
bosons or quarks; its production cross section at colliders would then be
small. However, in such a situation the bound of about 150 GeV also applies
to the next--to--lightest Higgs scalar. It can therefore be shown \cite{91}
that an \epem\ collider with $\sqrt{s} \geq 300$ GeV must find at least one of
the Higgs bosons of this model. However, as in the case of the MSSM this boson
might be very difficult to distinguish from the SM Higgs boson.

It has recently been argued by Sher \cite{92} that the running of the gauge
couplings might not even be changed significantly if $\lambda$ becomes
non--perturbative at some scale well below \mx. In this case {\em no} upper
bound on the Higgs boson mass can be given; indeed, just like in the heavy
Higgs limit of the SM, the existence of a recognizable (narrow) Higgs field
itself cannot strictly be guaranteed if Higgs self--interactions become
non--perturbative already at the TeV scale. However, it seems to us that
requiring the Higgs sector to be well behaved (perturbative) up to some
very high energies is much better motivated in the supersymmetric case than
without SUSY, since in the latter case the quantum behavior of the theory is
in any case problematic. Moreover, it is not clear to us whether the usual
proof \cite{3,12} for the cancellation of quadratic divergencies in SUSY
theories still holds in such a strongly coupled model.

Because the lightest Higgs boson of this model might be dominantly a singlet,
the usual LEP Higgs mass bounds do not apply. Similarly, the existence of a
fifth neutralino state can have consequences for collider phenomenology
\cite{94} as well as cosmology \cite{95}. Finally, we mention that this model
allows the charged Higgs boson to be lighter than the $W$ boson, in contrast
to the MSSM \cite{89}.

Since the singlet superfield does not affect the running of the gauge couplings
at the one--loop level, eqs.(\ref{e2.16})--(\ref{e2.20}) still hold. However,
the third generation sfermion masses, the masses of the heavy Higgs bosons, and
the effective $\mu-$parameter might all differ significantly from the MSSM. A
full RG study is needed to address this question quantitatively \cite{96}.
Unfortunately two recent analyses of this model reached somewhat contradictory
conclusions \cite{97,98}. No definite statement can therefore be made at
present, although we still expect the light stop and left--handed sbottom
squarks to be significantly lighter than first and second generation squarks;
the reduction of the masses of the other third generation scalars for large
values of \tanb\ should also remain qualitatively as in the MSSM, but
quantitative details will in general differ.

The motivation for an even more extensive modification of the Higgs sector of
the MSSM is in our opinion rather weak, unless it is required by the
introduction of a larger gauge group; see sec.~4c. We merely mention the
possibility that each generation of quarks and leptons might come with its
own pair of Higgs doublets \cite{99}. However, in such a model the suppression
of Higgs exchange contributions to FCNC processes is no longer automatic
\cite{100}. Moreover, even the first two generations of quarks and leptons
might now have sizable Yukawa couplings, if the corresponding vevs are very
small (or zero); this might lead to a very complicated (non--degenerate)
scalar spectrum at the weak scale.

It should be emphasized that, no matter how complicated the Higgs, gauge
and matter sectors are, one can still give quite a strong upper bound on the
mass of the lightest neutral Higgs boson, {\em if} we require all couplings
to remain perturbative up to scales of order \mx\ \cite{101}; in fact, this
bound is numerically very close to the bound of 150 GeV derived in the model
with additional singlet. Moreover, if this bound is saturated the lightest
Higgs boson is an $SU(2)$ doublet. We therefore expect that the ``no--lose"
theorem \cite{49,91} for SUSY Higgs searches at \epem\ colliders with
$\sqrt{s} \geq 300$ GeV (and $\int {\cal L} dt \geq 10$ fb$^{-1}$) holds in
{\em all} weakly coupled models.

\setcounter{footnote}{0}
\subsection*{4c) Models with Extended Gauge Sector}
Models with extended gauge sector at the TeV scale, i.e. additional charged
and/or neutral gauge bosons, were very popular in the mid--80's, when it
was believed that superstring models singled out the rank--6 GUT group $E(6)$;
see ref.\cite{102} for a review of the phenomenology of these $E(6)$ based
models. Here we are more interested in possible implications of an extended
gauge sector on the spectrum of sparticles that are already present in the
MSSM; after all, SUSY predicts that the superpartners of known SM particles
{\em must} exist, independent of possible extensions of the model.

We have already seen in sec.~3b that an enlargement of the gauge group will in
general introduce new $D-$term contributions to scalar masses. This is true
independent of the scale where the group breaks down to the SM gauge group; the
only condition is that there has to be some splitting between the masses of the
scalar fields responsible for this symmetry breaking \cite{75}. If this
symmetry breaking occurs significantly below \mx, the new gauge interactions
will also change the running of sfermion masses, i.e. the coefficients of the
terms $\propto m^2_{1/2}$ in eqs.(\ref{e2.16}) will change. This can be
especially significant for the $SU(3)_c \times SU(2)_L$ singlet fields
$\tilde{e}_R$, since the MSSM predicts their masses to run only slowly. If they
are embedded into $SU(2)_R$ doublets or even $SU(4)_c$ quartets their masses
will run much faster below \mx. In contrast, the relation (\ref{e2.18})
between gaugino masses still holds \cite{74} if the gauge group is unified
into a GUT at some scale.

The presence of new gauge bosons at the TeV scale means that the chargino,
neutralino and Higgs sectors also have to be extended. The masses of these new
fields are usually linked intimately to the masses of the new gauge bosons,
which are now known to be quite high \cite{31}; otherwise they would have been
detected directly at the Tevatron or indirectly (via $Z-Z'$ mixing) at
LEP/SLC. Indeed, the necessity to achieve $m^2_{Z',W'} \gg m^2_{Z,W}$ can be
considered to constitute a second hierarchy problem. If the Higgs bosons that
give masses to the new gauge bosons couple to those that break the \sym\
symmetry of the SM, the parameters of the Higgs potential have to be
fine--tuned to produce this hierarchy \cite{103}; note that such a coupling
will be present (through the $D-$terms) whenever the Higgs fields responsible
for \sym\ breaking transform nontrivially under the new gauge group, which is
true in all models where the SM gauge group is extended other than by simply
appending additional factor groups ($G = SU(3)_c \times \sym \times G'$).
While the necessary fine--tuning is much less severe than that needed to
produce the hierarchy between \mz\ and \mx, it cannot be protected by SUSY,
since the SUSY breaking scale in such models is of order $m_{Z'}$, {\em not}
of order \mz\ \cite{103,104,104a}.\footnote{This problem does not exist in
models where the Higgs potential possesses a $D-$flat direction and is
stabilized by non--renormalizable terms \cite{104b}. However, in in this case
the new gauge bosons typically have masses of order $10^9$ GeV or more and
thus completely decouple from the TeV scale; see \cite{104a} and references
therein. The sfermion spectrum will in general still differ from that of the
MSSM, however, as discussed above.} If a new $Z'$ boson were to be discovered
we  would therefore expect the SUSY breaking scalar masses to be quite high,
at least if we assume unification of these masses at some scale. In spite of
their large masses the new gauginos and higgsinos might then be produced in
the decays of MSSM squarks and sleptons. In such models the MSSM gauginos
could still be reasonably light if $\mhalf \ll m_0$. Another drawback of
models where the SM gauge group is embedded nontrivially in a larger group is
that the apparent unification of gauge couplings in the MSSM becomes
accidental.

In view of these problems the motivation for most $Z'$ models seems quite weak
to us. An exception might be models with an $SU(2)_R$ (or $U(1)_{B-L}$)
symmetry \cite{105}. The breaking of this symmetry can give rise to
non--vanishing neutrino masses, as might be required to explain the observed
solar and atmospheric neutrino anomalies. However, the elegant see--saw
mechanism \cite{106} for the generation of very small neutrino masses typically
predicts $SU(2)_R$ to be broken at scale $M_I \simeq m^2_{\mu}/m_{\nu_\mu}
\simeq 10^9$ GeV for $m_{\nu_\mu} \simeq 0.01$ eV as required for the standard
solution of the solar neutrino problem \cite{107}, making the new gauge bosons
and their superpartners completely unobservable.

\subsection*{4d) Models with Enlarged Matter Sector}
Additional matter superfields have also often been studied in the context of
``string--inspired" E(6) models \cite{102,104}. In addition to the fields of
the MSSM, a complete {\bf 27} of $E(6)$ contains vectorlike charge $= -1/3$
quarks\footnote{These fields might be di-- or lepto--quarks, depending on the
superpotential.} $g, \bar{g}$; vectorlike $SU(2)_L$ leptons $H, \bar{H}$; and
SM singlets $N$ and $\nu_R$. The most general superpotential compatible with
the low--energy gauge symmetry of the model contains quite a few terms, not all
of which are allowed to be present simultaneously if the proton is to be
sufficiently long--lived and neutrinos are to be sufficiently light. The usual
solution \cite{108} is to impose some discrete symmetries that constrain the
form of the superpotential. There are also weaker constraints on the size of
these couplings from rare processes \cite{109}. Many of these new couplings can
violate $R-$parity; see the following subsection. Moreover, couplings of the
type $L Q \bar{g}$ can greatly reduce the mass of $\tilde{L}$ sleptons, and
$\bar{U} \bar{D} \bar{g}$ couplings can lead to significant squark mass
splitting \cite{110}. Even in the absence of new superpotential couplings
eqs.(\ref{e2.26}) will have to be modified \cite{104}, since the presence of
new matter fields changes the $\beta-$functions for the gauge couplings.
Recall that we expect sfermions to be ``generically" quite heavy in such
models, since they contain additional gauge bosons.

The masses of the new vectorlike fermions are proportional to the vev of $N$
and can thus be quite large, of order of the $Z'$ boson mass. However, the
$\wt{N}$ fermions obtain masses only through mixing with $SU(2)_L$ doublets
$\wt{H}, \ \wt{\bar{H}}$ \cite{111}; they would be massless in the limit of
exact $SU(2)_L$ invariance. One can therefore derive an upper bound of about
120 GeV on the mass of the lightest neutral exotic lepton in such models
\cite{112}.

The motivation for such $E(6)$ type models now appears somewhat weak. A much
better case can be made for the existence of right--handed neutrino
superfields: They are necessary if a complete generation of the SM is to be
united in a single representation of a GUT group [the {\bf 16} of $SO(10)$],
and allow to produce nonvanishing neutrino masses. As gauge singlets they do
not affect the MSSM sparticle spectrum at all, unless one also enlarges the
gauge group (see the preceding subsection). However, the $\tilde{\nu}_R$ fields
might play a crucial role in creating the baryon density of the universe
\cite{113}.

\setcounter{footnote}{0}
\subsection*{4e) Models with Broken $R-$parity}
The MSSM as defined in sec.~2a is invariant under a discrete $Z_2$ symmetry
called $R-$parity. All SM quarks, leptons and gauge fields as well as the
Higgs bosons of the MSSM are even under this parity, whereas their
superpartners are odd. Conservation of $R-$parity implies that sparticles can
only be produced in pairs, and that the lightest sparticle is absolutely
stable. This obviously has immediate consequences for SUSY searches at
colliders \cite{53}.

However, the superpotential (\ref{e2.1}) is {\em not} the most general one
that is compatible with the gauge symmetries of the MSSM; rather, the following
terms can be added:
\be \label{e4.5}
W_R = \lambda_L L_L L_L E_R + \lambda'_L L_L Q_L D_R + \lambda_B U_R D_R D_R,
\ee
where generation indices have again been suppressed. Note that the terms
$\propto \lambda_L, \lambda'_L$ violate lepton number, while the term $\propto
\lambda_B$ violates baryon number; the simultaneous conservation of lepton and
baryon number always guarantees the existence of an exact
$R-$parity.\footnote{The inverse is not true, i.e. there might be an exact
$R-$parity even if baryon and/or lepton number are broken; this is the case,
e.g., in minimal SUSY $SU(5)$.} The bound on the proton lifetime then implies
upper bounds of order $10^{-25}$ or stronger on the products $\lambda_L \cdot
\lambda_B, \lambda'_L \cdot \lambda_B$, for sfermion masses of 1 TeV or less.
One therefore usually assumes either $\lambda_B = 0$ or $\lambda_L =
\lambda'_L = 0$. Even in that case bounds on the magnitude of certain
couplings in eq.(\ref{e4.5}) can be derived from bounds on rare processes like
$\mu \rightarrow e \gamma, \ K \rightarrow \pi l^+l^-$, etc. \cite{114}.

The existence of $R-$parity breaking terms in the superpotential can change
SUSY phenomenology quite dramatically. If some of the couplings in (\ref{e4.5})
are of order of the gauge couplings, they have to be included in all stages of
cascade decays of heavy sparticles; moreover, new single--sparticle production
processes can become important \cite{53}. In the presence of sizable $R-$parity
breaking couplings the running of the corresponding sfermion masses would be
altered, so that some of the eqs.(\ref{e2.16}) might no longer hold. Large
$R-$parity violating couplings involving third generation quarks would also
change the ``fixed--point" value for the top Yukawa coupling \cite{114a}; the
same analysis also gives upper bounds on these new couplings from the
requirement that they remain perturbative up to the GUT scale.

If the new couplings are much smaller than gauge couplings but larger than
about $3 \cdot 10^{-8} \left( \frac { 100 \ {\rm GeV} } { m_{\tilde{Z}_1} }
\right)^{1/2} \left( \frac { m_{\tilde{f}} } { m_{\tilde{Z}_1} } \right)^2$,
the
only effect of $R-$parity breaking is that the LSP (assumed to be \zino, as
favored by the RG analysis of sec.~2) will decay inside the detector. If the
$R-$parity couplings are even smaller than this, but larger than about $3 \cdot
10^{-15} \left( \frac { 100 \ {\rm GeV} } { m_{\tilde{Z}_1} } \right)^{1/2}
\left( \frac { m_{\tilde{f}} } { m_{\tilde{Z}_1} } \right)^2$, collider
phenomenology will not change, but \zino\ will decay sufficiently fast not to
endanger successful SM predictions for nucleosynthesis \cite{38}; recall that
both baryon and lepton number might be broken simultaneously if the couplings
in (\ref{e4.5}) are less than $10^{-13}$ or so.\footnote{The LSP decay width is
proportional to the square of some $R-$parity breaking coupling, while the
proton decay width is quartic in those couplings. One can therefore
simultaneously have $\tau_{\tilde{Z}_1} \leq 10^6$ sec and $\tau_p \geq
10^{31}$ yrs even if all couplings in (\ref{e4.5}) are of comparable size.} In
this case the LSP relic density constraint on the parameter space of the MSSM
(see sec.~2f) obviously no longer applies. Moreover, the LSP might now be
charged; in the MSSM with broken $R-$parity, the lighter stop eigenstate and
(if $m_0 \ll \mhalf$ and/or $\tanb \gg 1$) the lighter stau eigenstate are
possible candidates for a charged LSP. This will again change collider
phenomenology, especially if the LSP does not decay inside the detector.

Clearly SUSY phenomenology depends quite strongly on which, if any, of the
couplings in (\ref{e4.5}) are sizable, i.e. whether baryon or lepton number is
broken. From a theorist's perspective it seems attractive to ensure the
longevity of the proton by a (discrete) gauge symmetry, since otherwise the
symmetry is likely to be broken by quantum gravity effects. Viable gauge
symmetries, including discrete ones, are constrained by anomaly cancellation
conditions. In particular, it was shown in ref.\cite{115} that with the
particle content of the MSSM, only two viable discrete $Z_N$ symmetries
survive. One is the standard $R-$parity, and the other is a $Z_3$
symmetry that ensures baryon number conservation but allows lepton number
violation. This second symmetry actually forbids {\em all} proton decay,
because it implies a selection rule that baryon number can only be violated
in multiples of three units. More complicated anomaly--free discrete
symmetries also exist, including examples with conserved lepton number and
broken baryon number \cite{115}.

In any case, it should be clear that none of the terms in the superpotential
(\ref{e4.5}) need exist in a viable SUSY model. The principle of minimality
(``Occam's razor") then argues against their existence. Moreover, the
presence of a GUT group often strongly constrains or even completely excludes
the possibility of explicit $R-$parity violation. For instance, in minimal
$SU(5)$ all couplings in the superpotential (\ref{e4.5}) are proportional to
each other, and must therefore be of order $10^{-12}$ or less, as discussed
above. Moreover, $R-$parity conservation is actually automatic in extensions
with gauged $B-L$ as long the only nonvanishing VEVs are of fields with even
value of $3 \cdot (B-L)$ \cite{115a,115}; many $SO(10)$ models fall in this
category.

Even if the Lagrangian itself is $R-$parity invariant, it might still be
broken spontaneously. This possibility has mostly been discussed in the context
of supersymmetric Majoron models \cite{116}, where $R-$parity breaking is
linked intimately to the generation of neutrino masses. Such models predict
mixing between neutralinos and neutrinos, and between charginos and charged
leptons, where the upper bound on the mixing angles is related to the upper
bounds on neutrino masses. If $m_{\nu_{\tau}} \simeq 10$ MeV, these effects
might still be observable \cite{116}. In principle $R-$parity can even be
broken radiatively in the MSSM \cite{117} by nonzero (small) vevs of some
sneutrinos. However, in order to avoid the creation of an unacceptable
massless $SU(2)_L$ doublet Majoron one has to introduce some explicit
$R-$parity breaking as well, e.g. by adding a term $\mu' L_L H_2$ to the
superpotential. Finally we mention the result of Kuchimanchi and Mohapatra
\cite{105} that $R-$parity has to be broken spontaneously in supersymmetric
left--right symmetric models, by vevs of right--handed sneutrinos. In this
case the LSP might decay invisibly, leaving collider phenomenology unaltered.

\subsection*{4f) Models of Dynamical SUSY Breaking}
So far we have assumed that SUSY is broken in a ``hidden sector" by some
unknown dynamics; SUSY breaking is then transmitted to the visible sector by
interactions that are suppressed by some power of \mpl. In this approach the
only thing we need to know about SUSY breaking, at least as far as
phenomenological considerations are concerned, are the boundary conditions,
eq.(\ref{e2.3}) or their modifications discussed in secs.~3 and 4a. In this
subsection we summarize recent attempts \cite{118} to understand the dynamics
of
SUSY breaking in more detail. While these results could in principle also be
used in hidden sector models, the characteristic feature of this approach is
that it allows SUSY breaking to occur at much lower scales, which can have
experimentally testable ramifications.

In these models one introduces additional non--abelian gauge groups, which
break SUSY by instanton--effects if the right number of chiral superfields is
assumed \cite{119}. This is quite similar to the hidden sector of some
supergravity or superstring models. However, one also postulates the existence
of a ``messenger sector" that transmits SUSY breaking to the gauge and chiral
superfields of the MSSM. In ref.\cite{118} this sector consists of a pair of
vectorlike $SU(2)_L$ singlet quarks and a pair of vectorlike doublet leptons,
as well as several fields that are singlets under the SM gauge groups but
might carry the charge of a new $U(1)$ gauge group. Note that the MSSM matter
and Higgs fields are singlets under all the new gauge group factors; moreover,
the masses of the messenger fields are expected to lie in the 10 to 100 TeV
range. It will therefore be difficult to directly probe these new sectors
experimentally.

Nevertheless a number of testable predictions can be derived \cite{118}. To
begin with, gaugino masses fulfill the ``unification condition" (\ref{e2.18}),
i.e. are proportional to the squared gauge couplings. This is because they
are produced radiatively by diagrams involving the vectorlike messenger fields.
Sfermion masses are also produced radiatively. They are proportional to the
gaugino masses, with known coefficients. For example, $\msq \simeq \frac{16}
{\sqrt{3}} \mg$ (up to electroweak and higher loop corrections), and
$m_{\tilde{e}_R} \simeq 8 M_1$, where $M_1$ is the bino mass. Note that
sfermions with equal gauge quantum numbers are automatically degenerate in
this model, so that there are no problems with FCNC. The reduction of stop
(and $\tilde{b}_L$) masses should be less here than in the MSSM, since RG
scaling only occurs over $\sim 5$ (rather than $\sim 13$ or so) orders of
magnitude, and $A-$terms are expected to be small.

Since all masses are supposed to be created dynamically in this model, the
$\mu-$term has to be replaced by the vev of a singlet Higgs field, as described
in sec.~4b. Finally, since the SUSY breaking scale is relatively low here, the
gravitino is the lightest sparticle, with mass in the few keV range. The
lifetime for $\zino \rightarrow \wt{G} + \gamma$ decays is estimated to lie in
the range $10^{-13} - 10^{-5}$ seconds, most of which could be covered quite
easily in laboratory experiments once a SUSY signal has been found. This last
prediction should be very generic for this class of models, since it directly
follows from the low SUSY breaking scale; it is independent of the details of
the SUSY breaking and messenger sectors.

In the model of ref.\cite{118} the new fields that are nonsinglets under the
SM gauge group fill complete multiplets of $SU(5)$; they do therefore not
change the predicted value of \stw\ if Grand Unification of the SM gauge
interactions is assumed. However, if {\em all} gauge interactions of this
model were to be unified, a very large GUT group (with rank $\geq 8$) would
be needed. Moreover, in the standard hidden sector scenario the transmission
of SUSY breaking to the visible sector is basically automatic; it is caused
by terms in the Lagrangian whose presence is dictated by local supersymmetry,
although their exact form (and hence the sparticle spectrum) cannot be
predicted as yet. From this point of view the introduction of a messenger
sector, which is unavoidable if SUSY is to be broken at scales below
$\sqrt{\mpl
\mz}$, appears to be an unnecessary complication. On the other hand, in this
model the absence of FCNC can be proven rigorously; in the standard approach
it follows from certain {\em assumptions} about the scalar spectrum at very
high energies. Given that the model appears to be phenomenologically viable,
one
might as well go ahead and test its predictions.

\section*{5) Concluding Remarks}
A large number of SUSY models have been suggested in the literature; even the
subsample discussed in this article might leave some readers bewildered as
to what is a ``generic" SUSY prediction and what is contingent on details of
specific models.

To state the obvious, weak--scale supersymmetry uniquely predicts the
existence of superparticles ``at the weak scale". As discussed in sec.~2f,
even in the MSSM it is difficult to make this statement more quantitative.
As long as signals for sparticle production do not differ strongly from
those discussed in the framework of the MSSM, the failure to detect SUSY at
the LHC would certainly make the idea of weak--scale SUSY much less
attractive. It is at present not clear whether SUSY models can be constructed,
e.g. by explicitly breaking baryon number conservation and hence $R-$parity,
where squarks and gluinos would be essentially unobservable at hadron colliders
even if they are produced copiously; it is certainly fair to say that no
real motivation for this kind of SUSY model has yet been advanced. We are not
aware of any model where the visibility of charginos and sleptons at \epem\
colliders might be jeopardized; the failure to detect these sparticles
therefore
leads to almost completely model--independent bounds on their masses. However,
a 500 GeV collider, which is now being discussed together with the LHC, might
in
our view not be quite sufficient to decisively test the idea of weak--scale
SUSY.

We have emphasized repeatedly that the mass of the lightest neutral Higgs boson
has to be less than about 150 GeV in all weakly coupled SUSY models. However,
quite similar predictions can be made also in weakly coupled nonsupersymmetric
models. A light Higgs does appear to be more natural in the context of
supersymmetry, but discovery of the former is not (quite) sufficient to prove
the latter.

Turning to relations between sparticle masses, by far the most reliable
prediction is the sum rule (\ref{e3.1}) connecting masses of members of the
same $SU(2)_L$ sfermion doublet. The ``unification condition" (\ref{e2.18})
for gaugino masses also seems to be relatively robust; it even holds in some
models that are not unified at all, as discussed in sec.~4f. However, this
relation is not protected by a gauge symmetry. One can therefore construct
renormalizable (well--behaved) models where it is broken, although at present
the motivation for doing so is lacking.

We feel that the possibility of radiative gauge symmetry breaking described in
sec.~2b is a great advantage of SUSY models. However, it is difficult to
derive model--independent predictions from the assumption of radiative
symmetry breaking alone. In models with minimal particle content below the GUT
scale, i.e. in Grand Desert scenarios, one would generically expect a large
$\mu-$parameter, i.e. heavy higgsinos, given that the top is heavy. However,
if we allow some fine--tuning as well as non--degenerate scalar masses at
scale \mx, small values of $\mu$ cannot strictly be excluded.

We feel that models with automatic suppression of the most dangerous (gluino
induced) FCNC are preferable. However, as shown in sec.~4a, this does not
necessarily imply the degeneracy of scalars with equal gauge quantum numbers.
Assuming equal masses, say, for $\tilde{u}_R$ and $\tilde{d}_R$ type squarks
can at present only be motivated by simplicity arguments. In contrast, the
degeneracy of the eight gluinos is obviously guaranteed by $SU(3)_c$
invariance. Searches for gluino production at hadron colliders might therefore
be somewhat less model dependent than squark searches, although gluino decay
branching ratios, and hence signals for gluino production, do depend on the
squark spectrum.

The MSSM as described in sec.~2 is certainly a very attractive SUSY model. It
has a fairly small number of free parameters, which makes it quite predictive,
and easily passes all experimental tests. Unfortunately it seems quite
unlikely to us that it is exactly correct. For example, present data offer
tantalizing hints at a Grand Unified theory, see fig.~2; we saw in sec.~3b
that the presence of a GUT sector could modify some predictions of the MSSM
substantially. It should be admitted, however, that no really compelling SUSY
GUT has yet been suggested. The minimal $SU(5)$ model needs extreme
fine--tuning of parameters. Models that solve this problem tend to be either
quite complicated (like models employing the missing partner mechanism), or
offer only partial unification (like flipped $SU(5)$ models). From a
theoretical point of view the difficulty in understanding the apparent
unification of gauge couplings in the MSSM at a scale around $10^{16}$ GeV in
the framework of superstring theory, either with or without the presence of a
GUT sector, is also quite annoying. At present we can therefore only state
that even within the Grand Desert scenario we expect some modifications of the
MSSM due to physics at very high scales, $\geq 10^{16}$ GeV. These
modifications may only amount to a few percent, as in threshold corrections to
gauge couplings, but might also be much more significant. Of course, it is
also possible that the MSSM needs to be changed more extensively, see sec.~4.
One should therefore always be aware that testing the MSSM is {\em not} the
same as testing SUSY.

It is perhaps worth emphasizing again that even within the MSSM the
sparticle spectrum at the weak scale is more complicated than has often been
assumed. In particular, squarks and sleptons do usually {\em not} have the
same masses. Assuming a fixed mass ratio, independent of gaugino masses, is
hardly better. Moreover, at least some third generation squarks are expected
to be lighter than those of the first two generations. The failure to
discover these mass splittings would indicate a rather complicated form of
the spectrum at high energies, and/or new interactions involving light quarks.
Since $|\mu|$ is large, we expect most Higgs bosons to be quite heavy, unless
\tanb\ is large, in which case we expect the mass splitting between the first
two and the third generation to also exist in the slepton sector. In general it
seems to us that having a highly degenerate spectrum at very high energies,
where physics is supposed to be simple (i.e. unified) is preferable, even if
this implies a fairly diverse spectrum at experimentally accessible energies.

Lest the reader be discouraged we want to end on an upbeat note. At least
within supergravity (or string) models, studying sparticle masses and mixings
is {\em the} best way to learn about Planck scale physics. Even in models of
dynamical SUSY breaking discussed in sec.~4f the sparticle spectrum contains
information about physics at mass scales which will not be directly accessible
to experiment within our lifetime. Of course, finding a superparticle would be
a very great discovery in itself. Measuring sparticle masses and interactions
will then allow us a glimpse at truly fundamental physics.

\noindent
\subsection*{Acknowledgements}
We thank X. Tata, G.L. Kane and V. Barger for a critical reading of the
manuscript and useful comments, and Y. Okada for discussions on models with
extended Higgs sector. We owe special thanks to Heath Pois for his
collaboration during the early stages of this project, as well as a careful
reading of the final manuscript. The work of M.D. was supported in part by the
U.S. Department of Energy under grant No. DE-FG02-95ER40896, by the Wisconsin
Research Committee with funds granted by the Wisconsin Alumni Research
Foundation, as well as by a grant {}from the Deutsche Forschungsgemeinschaft
under the Heisenberg program. The work of S.P.M. was supported in part by the
US Department of Energy.

\end{document}